\documentstyle[epsf,12pt]{article}
\include{feynman}
\mark{{}{}}
\def\permil{\%\raise.10ex\hbox{$_{\scriptstyle 0}$}}

\begin{document}

\def\la{\mathrel{\mathpalette\fun <}}
\def\ga{\mathrel{\mathpalette\fun >}}
\def\fun#1#2{\lower3.6pt\vbox{\baselineskip0pt\lineskip.9pt
  \ialign{$\mathsurround=0pt#1\hfil##\hfil$\crcr#2\crcr\sim\crcr}}}
\begin{titlepage}
\centerline{\LARGE Gauge invariant effective action}
\vskip 0.4cm
\centerline{\LARGE  for high energy processes in QCD}

\vskip 0.4cm
\centerline{L. N. Lipatov $^{\dagger}$}
\vskip 0.3cm

\begin{center}
{
\it Petersburg Institute of Nuclear Physics,
Gatchina, 188350, St. Petersburg,  Russia }
\end{center}
\begin{center}
{
\it and}
\end{center}
\begin{center}
{
\it Deutsches Elektronen-Synchrotron DESY, Hamburg}
\end{center}
\vskip 15.0pt

\centerline{\bf Abstract}
\noindent
The Born amplitudes for a quasi-multi-Regge kinematics of produced gluons
are constructed in accordance with the Feynman rules
including apart from usual Yang-Mills vertices also an infinite number of
induced vertices. The new vertices describe the interaction of  physical
gluons
produced in  direct channels with the reggeized gluons propagating in
crossing channels.
The nonlinear gauge invariant effective action
reproducing these Feynman rules is constructed with the use of the Wilson
contour
integrals. After integrating over the physical degrees of freedom the
reggeon action is derived.
\vskip 3cm
\hrule
\vskip.3cm
\noindent

\noindent
$^{\dagger}${\it Humboldt Preistr\"ager\\
Work supported partly by INTAS and
the Russian Fund of Fundamental Investigations, grant No. 92-02-16809}
\vfill
\end{titlepage}

\section{Introduction}

The modern theoretical description of the deep inelastic $ep$ scattering at
small Bjorken variable $x=Q^2/(2m\nu)$ is based on the GLAP [1] and BFKL [2]
evolution equations. The GLAP equation allows one to predict the $Q^2$
dependence of the parton destributions $n_i (x)$ if they are measured at
a sufficiently large value of $Q=Q_0$ . In turn, the BFKL equation determines
their $x$ dependance in the small x range.

The most important distribution at small $x$
is the inclusive gluon density $g(x,k_\perp)$ which depends apart from
the longitudinal Sudakov component $x$ of the gluon momentum $k$
also on its transverse projection $k_\perp$ in the infinite momentum frame
of the proton. $g(x,k_\perp)$ is proportional to the
total gluon-proton scattering cross-section in the Regge regime
$s=-k_\perp^2/x \gg -k_\perp^2$ . In this region
the most probable process is the multi-gluon production.

In each order of perturbation theory
for gluon-gluon collisions at high c.m. energies $\sqrt{s}$
($s=2p_Ap_B $ )
the main contribution to the total cross-section $\sigma_{tot}$
results from the multi-Regge kinematics for final state
gluon momenta $k_0=p_{A'}, k_1,...,k_n, k_{n+1}=p_{B'}$ :

\begin{eqnarray}
s \gg s_{i}=2k_{i-1}k_i \gg t_i=q_i^2=(p_A-\sum\limits_{r=0}^{i-1}
k_r)^2,\,\,
\prod\limits_{i=1}^{n+1} s_i=s\prod\limits_{i=1}^n {\bf k}_{i}^2 \,\,,\,\,
k_{\perp}^2=-{\bf k}^2.
\end{eqnarray}

In the leading logarithmic approximation (LLA) the $n$-gluon production
amplitude in this kinematics has the multi-Regge form [2]:
\begin{eqnarray}
A_{2+n}^{LLA}=A_{2+n}^{tree} \prod\limits_{i=1}^{n+1} s_i^{\omega(t_i)}\,\,.
\end{eqnarray}
Here $s_i^{\omega(t_i)}$ are the Regge-factors appearing due to the known
fact [2], that gluons in each crossing channel $t_i$
are reggeized if one takes into account the radiative corrections to the
Born production amplitude $A_{2+n}^{tree}$. The gluon Regge trajectory in
LLA is $j=1+\omega(t)$, where
\begin{eqnarray}
\omega(t)=-\frac{g^2 N_c}{16\pi^3}\int d^2{\bf k}
\frac{{\bf q}^2}{{\bf k}^2 ({\bf q-k})^2} \,\,\,,\,\,t=-{\bf q}^2 \,\, .
\end{eqnarray}

The infrared divergencies in the Regge factors cancel in the total
cross section
with the contribution of the real gluons. The production amplitude
in the tree approximation has the following factorised form

\begin{eqnarray}
A_{2+n}^{tree}=2 s gT_{A'A}^{c_1}\Gamma_{A'A}\frac{1}{t_1}gT_{c_2 c_1}^{d_1}
\Gamma_{2,1}^1\frac{1}{t_2}....
gT_{c_{n+1}c_n}^{d_n}\Gamma_{n+1,n}^n
\frac{1}{t_{n+1}}gT_{B'B}^{c_{n+1}}\Gamma_{B'B}\,\,.
\end{eqnarray}
Here $A,B$ and $A',B',d_r $ ($r=1,2...n$) are colour indices for initial
and final gluons correspondingly.
$T_{ab}^{c}=-if_{abc}$ are generators of the gauge group
$SU(N_c)$ in the
self-conjugated representation, $g$ is the Yang-Mills coupling constant,

\begin{equation}
\Gamma_{A'A}=\delta_{\lambda_A,\lambda_A'}\,\, ,\,\,\,
\Gamma_{r+1,r}^{r}=C_{\mu}(q_{r+1},q_r)e_{\mu}^{\lambda_{r}}(k_r)
\end{equation}
are the reggeon-particle-particle (RPP) and
reggeon-reggeon-particle (RRP) vertices correspondingly; the quantity
$\lambda_r=\pm 1$ is the helicity of the gluon $r$ in the c.m.system.
In LLA the
$s$-channel helicities of colliding particles are conserved. Note that
in higher orders of perturbation theory for the RPP vertex this is not
the
case [3]. The effective RRP vertex $C(q_2,q_1)$ can be written as [2]
\begin{eqnarray}
C(q_2,q_1) = -q_1-q_2 +
p_A(\frac{q_1^2}{k_1p_A}+2\frac{k_1p_B}{p_Ap_B}) -
p_B(\frac{q_2^2}{k_1p_B}+2\frac{k_1p_A}{p_Ap_B}).
\end{eqnarray}
It has the important property corresponding to the current conservation
\begin{eqnarray}
(k_1)_\mu C_\mu (q_2,q_1)=0,
\end{eqnarray}
which gives us a possibility to chose an arbitrary gauge for each of the
produced gluons. For example, in the left ($l$) light cone gauges, where
$p_A e^l(k)=0$ and $k e^l(k)=0$, one can use the following
parametrisation of the polarization vector $e^l(k)$
\begin{eqnarray}
e^l=e_{\perp}^l-\frac{k_{\perp}e_{\perp}^l}{kp_A} p_A
\end{eqnarray}
in terms of the two dimensional vector $e_{\perp}^l$. In this gauge the
RRP vertex takes an especially simple form, if we introduce the complex
components $e=e_x+ie_y\, , e^*=e_x-ie_y$ and $k=k_x+ik_y\, , k^*=k_x-ik_y$
for transverse vectors $e_\perp^l ,k_\perp$ [4]
\begin{eqnarray*}
\Gamma_{2,1}^1=C e^* + C^* e ,\,\,\, C=\frac{q_1^* q_2}{k_1^*}.
\end{eqnarray*}
This representation was used in [4] to construct an effective scalar field
theory for the multi-Regge processes. The Lagrangian of this theory
includes apart from the effective RRP vertex $C$ also the RPP vertex
$\Gamma_{1,2}$ and the helicity conservation is reformulated here
as the charge conservation. The effective action  was
derived
recently from the Yang-Mills theory by integrating over the
fields which correspond to the  highly virtual particles produced in the
multi-Regge kinematics in the direct channels $s_i$ [5]. As a result of
this integration the induced terms appear in the Lagrangian.

The action for the effective field theory [4] describing multi-Regge
processes can be written in a form which is invariant under the abelian gauge
transformations $\delta V_\mu^a =i\partial_\mu \chi^a$ for the fields $V_\mu$
describing the physical gluons provided that the fields $A_\pm$
corresponding to the reggeized gluons  are gauge invariant ( $\delta
A_\pm =0$):
\begin{eqnarray*}
S_{m\,R}=\int d^4x\{\frac{1}{4}(F_{\mu \nu}^{a})^2+\frac{1}{2}
(\partial_{\perp \sigma}A_+^a)(\partial_{\perp \sigma}A_-^a)+
\end{eqnarray*}
\begin{eqnarray*}
+\frac{1}{2}\,g\,[-A_+^a(F_{-\sigma}
T^a\,i\partial_-^{-1}F_{-\sigma})-A_-^a(F_{+\sigma}T^a\,
i\partial_+^{-1}F_{+\sigma})+
(\partial_-^{-1}
F_{-\sigma}^a)(A_-T^a\,i\partial_\sigma A_+)+
\end{eqnarray*}
\begin{eqnarray}
+(\partial_+^{-1}F_{+\sigma}^a)(A_+T^a\,i
\partial_\sigma A_-)+i(\frac{1}{\partial_+}\frac{1}{\partial_-}F_{+-}^a)
(\partial_\sigma A_+)T^a(\partial_\sigma
A_-)+iF_{+-}^a(A_-T^aA_+)]\}\,,
\end{eqnarray}
where $F_{\mu \nu}=\partial_\mu V_\nu -\partial_\nu V_\mu$ and $N_\pm
=N_0\pm N_3$ are the light-cone components of the vectors $V_\mu$ in the
c.m. system ($p_A^+=p_B^-=\sqrt{s}$). In the interaction part of action (9)
two first terms
correspond to the gluon scattering vertices and four last terms lead to
the production vertex (6). In these contributions one can use
the equations $\partial_\pm A_\mp =0$ valid in the Regge kinematics.
The scalar field version [4] of the effective theory for the multi-Regge
processes can be obtained from action (9) if one will use the light cone gauge
(8) to eliminate from $V$ non-physical degrees of freedom. To avoid the
double counting of contributions of the Feynman diagrams one should remember,
that the induced terms in the effective vertices appear
as a result of the integration over momenta of highly virtual particles
in the direct channels [5].
Below we shall
construct a more general effective action invariant under nonabelian gauge
transformations.

The total cross-section
calculated in LLA using the above expressions for production amplitudes
grows very rapidly as $s^\omega$
($\omega=(g^2 N_c/\pi^2) \ln 2 $) and
violates the Froissart bound $\sigma_{tot} < c \ln^2 s$ [2]. One of the
possible
ways of the unitarisation of the LLA results is to use the above effective
field
theory [4,5] .

Another more simple
(but not so general) method is related to the solution of
the BKP equation [6] for multi-gluon compound states. This equation has
a number of remarkable properties, including  conformal invariance [7],
holomorphic separability [8], and the existence of nontrivial integrals of
motion [9]. The Hamiltonian for the corresponding Schr\"odinger equation
coincides with the Hamiltonian for a completely integrable Heisenberg model
with spins belonging to an infinite dimensional representation of the
noncompact M\"obius group [10]. It means, that the Bethe anzatz
can be used for finding
eigen values and eigen functions of this Schr\"odinger equation. Faddeev
and Korchemsky have shown [10], that for this purpose one should find the
solution of the Baxter equation for an integer function $Q(\lambda)$.

For two gluons the solution is known [10].
In the case $n\,>\,2$ one can present
$Q(\lambda)$ as a linear combination of these solutions, which leads to a
reccurence equation for its
coefficients $d_k$. For example, for $n=3$ this equation coincides with
the fundamental relation for the orthogonal polynomials $d_k(A)$
of the
discrete variable $A$ being the eigen value of the quantum integral
of motion obtained
in ref.[9]:
\begin{eqnarray*}
A\,d_k(A)=\frac{k(k+1)(k-m+1)(k+m)}{2(2k+1)}(d_{k+1}(A)\,+\,d_{k-1}(A)).
\end{eqnarray*}
Here $d_0=0,d_1=1$ and the quantization condition for $A$ at
integer values of the conformal weight $m$ is $d_{m-1}(A)=0$
(for $A \not= 0$).

The orthogonality and completeness
conditions for these polynomials can be constructed:
\begin{eqnarray*}
\sum_{A \not= 0}
\frac{d_k(A)\,d_{k_1}(A)}{d_{m-2}(A)\,d'_{m-1}(A)}=\delta_{kk_1}
\frac{k(k+1)(k-m+1)(k+m)}{2(2k+1)},
\end{eqnarray*}
\begin{eqnarray*}
\sum_{k=1}^{m-2}\frac{2(2k+1)\,d_k(A)\,d_k(A_1)}{k(k+1)(k-m+1)(k+m)}=
\delta_{AA_1}\,d_{m-2}(A)\,d'_{m-1}(A)\,,\,A \not= 0\,.
\end{eqnarray*}
Nevertheless, their theory is not developed enough
to find analitically
the wave function and the intercept of the perturbative
Odderon.

All these results are based on  calculations of  effective Reggeon
vertices
and the gluon Regge trajectory in the first nontrivial order of perturbation
theory. Up to now we  do not know the region of applicability of LLA
including
the low boundary for the initial energy and
the momentum scale for the  QCD coupling constant. Note, however, that
recently the  Born production amplitudes
for a quasi-multi-Regge kinematics of final particles
were calculated [11] and one-loop corrections to the
reggeon vertices were found [3], which will allow one to
find next to leading corrections to the BFKL equation.

In this paper we want to generalize the effective field theory approach
of ref. [4] to processes for which the final state particles are
separated in several groups consisting of
an arbitrary number of gluons with a fixed
invariant mass;  each group is produced with respect to others
in the multi-Regge
kinematics. These conditions are more general than the requirements for
the quasi-multi-Regge
kinematics of Ref. [11] where only one additional group consisting of two
gluons was considered.
We show  that the effective action for such a generalised quasi-multi-Regge
process can be written in a gauge invariant form in terms of the Wilson
contour integrals. (The Wilson contour integrals were used
by Ya. Balitsky in his talk at the Fermi Lab Small-x Meeting (September
1994) to
justify the application of the operator product expansion to the Regge
processes in QCD.)

\section{Quasi-elastic processes}

The gluon-gluon elastic scattering amplitude in the Born approximation
can be written as follows (see [2], cf.(4)):
\begin{equation}
A_2^{tree}=\frac{1}{2}\, g\, T_{A'A}^c\,
\Gamma^{\nu' \nu +}(p_{A'},p_A)\, e_{\nu' }^*\, e_{\nu }\, \frac{1}{t}\,
g\, T_{B'B}^c\, \Gamma^{\mu' \mu -}(p_{B'},p_B)\, e_{\mu' }^*\, e_{\mu }\,,
\end{equation}
where the tensor $\Gamma^{\nu_A'\nu_A +}$ contains apart from
the light-cone projection of the Yang-Mills (YM) vertex
\begin{eqnarray}
\gamma^{\nu'\nu \sigma}(p_{A'},p_A) = \delta^{\nu'\nu }\,
(p_{A'}+p_A)^\sigma +
\delta^{\nu \sigma }\, (-2p_A+p_{A'})^{\nu'} + \delta^{\nu'\sigma }\,
(-2p_{A'}+p_A)^{\nu }
\end{eqnarray}
an additional induced term (cf.[4,5]):

\begin{eqnarray}
\Gamma^{{\nu}'\nu +}(p_{A'},p_A) = \gamma^{{\nu}'\nu + }(p_{A'},p_A) -
t\,\, n^{+{\nu}'}\,\, \frac{1}{p_{A}^+}\,\, n^{+\nu} .
\end{eqnarray}
Here the light-cone vectors $n^+\,,\,n^-$ and the light-cone components
$k^+\,,\, k^-$ for particle momenta are defined in the c.m. system
($p_A^+=p_B^-=2E\,,s=4 E^2\,,p_A^-=p_B^+=0$) by the equations
\begin{eqnarray}
n^+=\frac{p_B}{E}\,,\, n^-=\frac{p_A}{E}\,,\,k^+=kn^+\,,\,k^-=kn^-\,,
\end{eqnarray}
where $n^+n^-=n^+n_+=2$ and $\partial_+x^+=2$. These definitions differ
slightly from
notations used in [4,5]. Using the Ward identity for the YM vertex
\begin{eqnarray}
(p_{A'})_{{\nu}'}\,\gamma^{{\nu}'\nu \sigma}(p_{A'},p_A) =
(t-p_A^2)\delta^{\nu
\sigma} + p_{A'}^\nu q^\sigma + p_{A'}^\sigma p_A^\nu
\end{eqnarray}
we obtain the corresponding identity
for the effective vertex $\Gamma^{{\nu}'\nu +}$
\begin{eqnarray}
(p_{A'})_{ {\nu}'}\Gamma^{{\nu}'\nu +}(p_{A'},p_A)= -p_A^2 \, (n^+)^\nu +
p_A^+ (p_A)^\nu,
\end{eqnarray}
where we neglected the small terms of the order of $(p_Bq)/(p_Bp_A)$.
If the gluon $A$ is on  mass shell
($p_A^2=0$) and  its polarization vector $e(p_A)$ satifies the
Lorentz condition $p_Ae=0$, we conclude from (15) that
the effective vertex is gauge invariant $(p_{A'})_{{\nu}'}
\Gamma^{{\nu}'\nu +}=0$. This means that
one can use an arbitrary gauge for the polarization vectors $e$.

In the light cone gauge $p_Be(p_A)=p_Be(p_{A'})=0$ we obtain
\begin{eqnarray}
e_{{\nu}'}^*(p_{A'})e_\nu (p_A) \Gamma^{{\nu}'\nu +}=-4E({\bf e}_{A'}^*{\bf
e}_A)=-4E\delta_{\lambda_{A'}\lambda_A}\,\,.
\end{eqnarray}
Therefore taking into account relations (5) one can verify
that (10) coincides with  (4) for the elastic case $n=0$.

Let us consider now the quasi-elastic process in which the final state
contains apart from the particle $B'$ with  momentum $p_{B'} \simeq p_B$
also several gluons with a fixed
invariant mass in the fragmentation
region of the initial gluon $A$. It is convenient to denote the colour
indices of the produced gluons by $a_1,a_2,...a_n$ leaving the index $a_0$
for the particle $A$. Further, the momenta of the produced gluons and of the
particle $A$ are denoted by $k_1,k_2,...k_n$, and $-k_0$ correspondingly.
$q=-\sum\limits_{i=0}^n k_i$ is the momentum transfer.  Omitting the
polarization vectors
$e_{{\nu}_i}(k_i)$ for the gluons $i=0,1,...n$ we can write the production
amplitude related with the single gluon exchange in the tensor representation
\begin{eqnarray}
A_{a_0a_1...a_nB'B}^{{\nu}_0{\nu}_1...{\nu}_n}\,=\,\,
-\phi_{a_0a_1...a_nc}^{\nu_0\nu_1...\nu_n  +}\,
\,\,\frac{1}{t}\,\,g\,\, p_B^- T_{B'B}^c
\delta_{\lambda_{B'},\lambda_{B}}\,.
\end{eqnarray}
Here the form-factor $\phi$ depends on the invariants constructed from the
momenta $k_0,..k_n$ .

For the simplest case of one gluon production $\phi$ was calculated  in the
Born approximation in [11]. We present this result in the form (cf.
Appendix)

\begin{eqnarray*}
{\phi}_{a_0a_1a_2c}^{{\nu}_0{\nu}_1{\nu}_2+} =
g^2\{\Gamma_{a_0a_1a_2c}^{{\nu}_0{\nu}_1{\nu}_2
+}-T_{a_1a_0}^aT_{a_2a}^c
\frac{\gamma^{{\nu}_1{\nu}_0\sigma }(k_1,-k_0)\Gamma^{{\nu}_2\sigma
+}(k_2,k_2+q)}{(k_0+k_1)^2}\,\,-
\end{eqnarray*}
\begin{eqnarray*}
-T_{a_2a_0}^aT_{a_1a}^c\frac{\gamma^{{\nu}_2{\nu}_0\sigma
}(k_2,-k_0)\Gamma^{{\nu}_1\sigma +}(k_1,k_1+q)}{(k_0+k_2)^2}-
\end{eqnarray*}
\begin{eqnarray}
-T_{a_2a_1}^aT_{a_0
a}^c\frac{\gamma^{{\nu}_2{\nu}_1 \sigma }(k_2,-k_1)\Gamma^{{\nu}_0\sigma
+}(k_0,k_0+q)}{(k_1+k_2)^2}\}.
\end{eqnarray}

The last three terms in the brackets correspond to
the Feynman diagram contributions constructed from the gluon
propagator combining the usual
Yang-Mills vertex (11) and the effective RPP vertex (12). The first term
can be written as
\begin{eqnarray}
\Gamma_{a_0a_1a_2c}^{{\nu}_0{\nu}_1{\nu}_2
+}\,=\,\gamma_{a_0a_1a_2c}^{\nu_0\nu_1\nu_2+}\,+\,\Delta_{a_0a_1a_2c}^
{\nu_0\nu_1\nu_2+}\,\,,
\end{eqnarray}
where $\gamma$ is the light-cone projection of
the usual quadri-linear Yang-Mills vertex
\begin{eqnarray*}
\gamma_{a_0a_1a_2c}^{\nu_0\nu_1\nu_2+}=T_{a_1a_0}^aT_{a_2a}^c(
\delta^{\nu_1\nu_2}\delta^{\nu_0+}-
\delta^{\nu_1+}\delta^{\nu_0\nu_2})+
\end{eqnarray*}
\begin{eqnarray}
+T_{a_2a_0}^aT_{a_1a}^c(\delta^{\nu_2\nu_1}
\delta^{\nu_0+}-\delta^{\nu_2+}\delta^{\nu_0\nu_1})+T_{a_2a_1}^aT_{a_0a}^c
(\delta^{\nu_2\nu_0}\delta^{\nu_1+}-\delta^{\nu_2+}\delta^{\nu_1\nu_0})
\end{eqnarray}
and $\Delta$ is a new induced vertex
\begin{eqnarray}
\Delta_{a_0a_1a_2c}^{\nu_0\nu_1\nu_2 +}(k_0^+,k_1^+,k_2^+)\,
=-t\,(n^+)^{\nu_0}
(n^+)^{\nu_1}(n^+)^{\nu_2}\{\frac{T_{a_2a_0}^aT_{a_1a}^c}
{k_1^+\,\,\,k_2^+}+\frac{T_{a_2a_1}^aT_{a_0a}^c}{k_0^+\,\,\,k_2^+}\}.
\end{eqnarray}
Note that due to the Jacobi identity

\begin{equation}
T_{a_2a_0}^aT_{a_1a}^c-T_{a_2a_1}^aT_{a_0a}^c=T_{a_1a_0}^aT_{a_2a}^c
\end{equation}
and the momentum conservation law

\begin{equation}
k_0^+\,+\,k_1^+\,+\,k_2^+\,=\,0
\end{equation}
which is valid in the quasi-elastic kinematics at large $s$
the tensor $\Delta$ is
Bose-symmetric with respect to the simultaneous transmutation of momenta,
colour and Lorentz indices of the gluons $0,1,2$.

Using the Ward identities (14) and (15) for vertices $\gamma$ and $\Gamma$
one can verify that for $n=2$ the amplitude (17)  is gauge invariant
\begin{eqnarray}
(k_i)_{\nu_i}A^{\nu_0\nu_1\nu_2}\,=\,0 \,\,\,\,,i=0,1,2 \, .
\end{eqnarray}

Let us include the colour matrices in the trilinear YM vertices
(see (11),(12)):
\begin{eqnarray}
\gamma_{a_0a_1c}^{\nu_0\nu_1\sigma}(k_1,-k_0)=
T_{a_1a_0}^c\gamma^{\nu_1\nu_0\sigma}(k_1,-k_0)\,\,,\,\,
\Gamma_{a_0a_1c}^{\nu_0\nu_1+}(k_1,-k_0)=
T_{a_1a_0}^c\Gamma^{\nu_1\nu_0+}(k_1,-k_0).
\end{eqnarray}
Then we obtain similarly to (19)

\begin{eqnarray}
\Gamma_{a_0a_1c}^{\nu_0\nu_1+}=
\gamma_{a_0a_1c}^{\nu_0\nu_1+}+\Delta_{a_0a_1c}^{\nu_0\nu_1+},
\end{eqnarray}
where, according to (12),
we have
\begin{eqnarray}
\Delta_{a_0a_1c}^{\nu_0\nu_1+}(k_0^+,k_1^+)\,=\,-tT_{a_1a_0}^c(n^+)^{\nu_1}
\frac{1}{k_1^+}(n^+)^{\nu_0}\,\,\,\, ,k_0^++k_1^+=0.
\end{eqnarray}

In the general case  $n>2$ for the gauge invariance of $\phi$  (17)
one should take into account apart from the usual YM vertices
$\gamma_3$, $\gamma_4$ and corresponding effective vertices (26,19)  also
the effective vertices $\Gamma$ coinciding with the induced vertices
$\Delta$ for an arbitrary number $r>3$ of external legs
\begin{eqnarray}
\Gamma_{a_0a_1...a_{r-1}c}^{\nu_0\nu_1...\nu_{r-1}+}(k_0^+,k_1^+,...k_{r-1}^+)=
\Delta_{a_0a_1...a_{r-1}c}^{\nu_0\nu_1...\nu_{r-1}+}(k_0^+,k_1^+,...k_{r-1}^+)
\,\,\,\,\,.
\end{eqnarray}
Let us consider the contribution of all Feynman diagrams constructed
from such induced vertex combined by a gluon propagator with the usual
YM vertex $\gamma$ (25) describing the interaction with an external gluon
having the momentum $k_r$ (cf. (18)).
Multiplying the corresponding expression for $\phi$
in eq. (17) by $k_r^{\nu_r}$ and
using the Ward identity (14) for $\gamma$ we obtain the non-vanishing sum
of terms:
\begin{eqnarray}
-\sum\limits_{i=0}^{r-1}
T_{a_ra_i}^a\Delta_{a_0a_1...a_{i-1}aa_{i+1}...a_{r-1}c}^{\nu_0...\nu_{r-1}+}
(k_0^+,...,k_{i-1}^+,k_i^++k_r^+,...,k_{r-1}^+).
\end{eqnarray}
To compensate it by  higher order contributions the infinite set of the
quantities $\Delta$ should satisfy the reccurence relation:
\begin{eqnarray*}
\Delta_{a_0a_1...a_rc}^{\nu_0\nu_1...\nu_r+}(k_0^+,k_1^+,...,k_r^+)=
\end{eqnarray*}
\begin{eqnarray}
\frac{(n^+)^{\nu_r}}{k_r^+}
\sum\limits_{i=0}^{r-1}T_{a_ra_i}^a
\Delta_{a_0 a_1...a_{i-1} a a_{i+1}...a_{r-1}c}^{\nu_0...\nu_{r-1}+}
(k_0^+,...,k_{i-1}^+,k_i^++k_r^+,k_{i+1}^+,...,k_{r-1}^+)\,.
\end{eqnarray}
These induced vertices are invariant under  arbitrary transmutations of
indices {i} :
\begin{eqnarray}
\Delta_{a_{i_0}a_{i_1}...a_{i_r}c}^{\nu_{i_0}\nu_{i_1}...\nu_{i_r}+}
(k_{i_0}^+,k_{i_1}^+,...k_{i_r}^+)=
\Delta_{a_0a_1...a_rc}^{\nu_0\nu_1...\nu_r+}(k_0^+,k_1^+,...k_r^+)
\end{eqnarray}
due to the Jacobi identity (22) for colour group generators $T$ and the
energy-momentum conservation
\begin{eqnarray}
\sum\limits_{i=0}^r k_i^+\,\,=\,\,0.
\end{eqnarray}
By constructing $\phi$ in (17) according to the Feynman rules including
apart from the usual Yang-Mills vertices $\gamma$ also the effective
vertices $\Gamma$  we obtain for it a gauge invariant expression.

\section{Multi-gluon production in the central region}
In the previous section we considered  a generalization of the
effective vertices $\Gamma_{A'A}^c=T_{A'A}^c\Gamma_{A'A}$ in eq.
(4) to the case of the
quasi-elastic processes. Here we want to generalize the effective vertices
$\Gamma_{c_2c_1}^{d_1}=T_{c_2c_1}^{d_1}\Gamma_{2,1}^1$ for the multi-gluon
production. For
simplicity we discuss the following kinematics of the final state particles:
the gluons $A'$ and $B'$ move almost along the momenta of
the initial particles $A$ and $B$
and there is a group of produced gluons with a fixed invariant mass in
the central region $y \sim 0$ of the rapidity $y=\frac{1}{2}\ln (k^+/k^-)$.
The momentum
transfers $q_1=p_A-p_{A'}$ and $q_2=p_{B'}-p_{B}$ in this regime have
(with   a good accuracy) the Sudakov decomposition
\begin{eqnarray}
q_1=q_{1\perp}+\beta \,p_A\,,\,\,q_2=q_{2\perp}-\alpha \,p_B\,
\end{eqnarray}
where $\beta$ and $\alpha$ are the Sudakov parameters of
the total momentum
$k=\sum\limits_{i=1}^n k_i$ of the produced gluons:
\begin{eqnarray}
k=k_\perp + \beta \,p_A + \alpha \,p_B\,\, ,\,\, \kappa =k^2=s\alpha \beta +
(q_1-q_2)_\perp^2
\end{eqnarray}
and $\sqrt{\kappa}$ is their invariant mass  which is asumed to be fixed
at high energies: $\kappa \ll s$.

In this kinematical region the production amplitude
has the factorized form  (cf. (4) and (17))

\begin{eqnarray}
A_{d_1d_2....d_n A'A B'B}^{\nu_1\nu_2...\nu_n+-}=-\,g\,p_A^+\,T_{A'A}^{c_1}
\,\Gamma_{AA'}\,\frac{1}{t_1}\,\psi_{d_1d_2...d_nc_2c_1}^{\nu_1\nu_2...\nu_n+-}
\,\frac{1}{t_2}\,g\,p_B^-\,T_{B'B}^{c_2}\, \Gamma_{B'B}.
\end{eqnarray}
For the case of  one gluon emission we have from eq.(6)
\begin{eqnarray}
\psi_{d_1c_2c_1}^{\nu_1+-}=g\,\Gamma_{d_1c_2c_1}^{\nu_1+-},
\end{eqnarray}
where $\Gamma$ is the sum of two terms
\begin{eqnarray}
\Gamma_{d_1c_2c_1}^{\nu_1+-}=\gamma_{d_1c_2c_1}^{\nu_1+-}+\Delta_{d_1c_2c_1}^
{\nu_1+-}.
\end{eqnarray}
The first term is the contribution from the tri-linear Yang-Mills vertex
(see (11))
\begin{eqnarray}
\gamma_{d_1c_2c_1}^{\nu_1+-}=T_{c_2c_1}^{d_1}\gamma^{\nu_1+-}\,,\,
\gamma = 2(q_2+q_1)-2k_1^+n^-+2k_1^-n^+.
\end{eqnarray}
The second term is the induced one
\begin{eqnarray}
\Delta_{d_1c_2c_1}^{\nu_1+-}=T_{c_2c_1}^{d_1}\Delta^{\nu_1+-}\,,\,
\Delta = -2\,t_1\,\frac{n^-}{k_1^-}\,+\,2\,t_2\,\frac{n^+}{k_1^+}\,\,.
\end{eqnarray}
Due to the relation (see (6))
\begin{eqnarray}
\gamma + \Delta =-2\,C,
\end{eqnarray}
expression (35) coincides with
(4) for the particular case $n=1$.

Let us consider now the production of two gluons in the central region. The
amplitude of this process in the Born approximation
was calculated in ref. [11]. The result can be
written in the form of representation (35) where the tensor $\psi$ is
(see Appendix):
\begin{eqnarray*}
\psi_{d_1d_2c_2c_1}^{\nu_1\nu_2+-}=g^2\{\Gamma_{d_1d_2c_2c_1}^{\nu_1\nu_2+-}
-\frac{T_{d_2d_1}^d
\gamma^{\nu_2\nu_1\sigma}
(k_2,-k_1)\,\,\Gamma_{d c_2c_1}^{\sigma +-}(q_2,q_1)}{(k_1+k_2)^2}\,\,\,\,-
\end{eqnarray*}
\begin{eqnarray}
-\frac{\Gamma_{d_1d c_1}^{\nu_1\sigma -}(k_1,k_1-q_1)
\Gamma_{d_2d c_2}^{\nu_2 \sigma +}(k_2,k_2+q_2)}{(q_1-k_1)^2}-
\frac{\Gamma_{d_2d c_1}^{\nu_2 \sigma -}(k_2,k_2-q_1)
\Gamma_{d_1d c_2}^{\nu_1 \sigma +}(k_1,k_1+q_2)}{(q_1-k_2)^2}
\}.
\end{eqnarray}

The second term in the brackets describes the production of a pair of gluons
through the decay of the virtual gluon in the direct channel. This
contribution is a product of the effective vertex $\Gamma$, the usual
YM vertex $\gamma$ and the gluon propagator. Analogously, the third and
fourth contributions are products of two effective vertices (25) having
the light cone components $\pm$ and the gluon propagator in the
crossing channels.

The first term in the brackets is not  expressed
in terms of the effective vertices which appear in LLA .  It can be
presented as the sum of two terms
\begin{eqnarray}
\Gamma_{d_1d_2c_2c_1}^{\nu_1\nu_2+-}=\gamma_{d_1d_2c_2c_1}^{\nu_1\nu_2+-}+
\Delta_{d_1d_2c_2c_1}^{\nu_1\nu_2+-},
\end{eqnarray}
where the contribution $\gamma$ is the light cone component of the
quadri-linear Yang-Mills vertex (cf.(20))
\begin{eqnarray*}
\gamma_{d_1d_2c_2c_1}^{\nu_1\nu_2+-}=T_{d_1c_1}^d T_{d_2d}^{c_2}
(\delta^{\nu_1\nu_2}\delta^{-+}-\delta^{\nu_1+}\delta^{-\nu_2})\,\,\,+
\end{eqnarray*}
\begin{eqnarray}
+\,\,T_{d_2c_1}^d
T_{d_1d}^{c_2}(\delta^{\nu_2\nu_1}\delta^{-+}-\delta^{\nu_2+}
\delta^{-\nu_1})+T_{d_2d_1}^dT_{c_1d}^{c_2} (\delta^{\nu_2-}\delta^{\nu_1+}-
\delta^{\nu_2+}\delta^{\nu_1-})
\end{eqnarray}
and $\Delta$ is the new induced vertex (cf. (21))

\begin{eqnarray*}
\Delta_{d_1d_2c_2c_1}^{\nu_1\nu_2+-}=-2\,t_2\, (n^+)^{\nu_1}(n^+)^{\nu_2}
\{\frac{T_{d_2c_1}^dT_{d_1d}^{c_2}}{k_1^+\,\,k_2^+}+
\frac{T_{d_2d_1}^dT_{c_1d}^{c_2}}{(-k_1^+-k_2^+)\,k_2^+}\}\,\,\,-
\end{eqnarray*}
\begin{eqnarray}
-2\,t_1\,(n^-)^{\nu_1}(n^-)^{\nu_2}\{\frac{T_{d_1c_2}^dT_{d_2d}^{c_1}}
{k_2^-\,\,k_1^-}+\frac{T_{d_1d_2}^dT_{c_2d}^{c_1}}{(-k_1^--k_2^-)\,k_1^-}
\}\,.
\end{eqnarray}
One can verify that $\Delta$ is Bose invariant and symmetric with
respect to the simultaneous transmutation of $n^-,t_1,d_1$ and
$n^+,t_2,d_2$.

For the gauge invariance of the production amplitude in the case $n>2$
one should introduce the induced vertices with an arbitrary number of
external legs which are expressed in terms of the light cone projections
of the vertices (28)
\begin{eqnarray}
\Gamma_{d_1...d_nc_2c_1}^{\nu_1...\nu_n+-}=\Delta_{c_1d_1...d_nc_2}^
{+\nu_1...\nu_n-}(k_0^+k_1^+...k_n^+)+\Delta_{c_2d_1...d_nc_1}^
{+\nu_1...\nu_n-}(k_0^-k_1^-...k_n^-)\,,
\end{eqnarray}
where $k_0^+,k_0^-$ are determined by the momentum conservation
\begin{eqnarray}
\sum\limits_{i=0}^n k_i^+\,=\,\sum\limits_{i=0}^n k_i^-\,=\,0.
\end{eqnarray}

In the next sections we consider the gauge-invariant
functional formulation of the
effective field theory for a general kinematics of the final state particles
in high energy processes.

\section{Field description of particles and reggeons in the Yang-Mills
theory}
Due to the gluon reggeization it is natural to expect  that QCD
at high energies   can be reformulated as an interaction theory for
physical particles (quarks and gluons) and reggeized gluons. Futhermore,
after integration over physical
degrees of freedom one should develop the Reggeon field theory analogous
to the Pomeron calculus which was invented  by V. Gribov many years ago
[12]. In comparison with the previous publications [4,5]
devoted to the multi-Regge processes in this paper
we attemp to construct the
Reggeon calculus for the gluodynamics in a form being similar to
one of ref.
[12]  starting from the action for physical gluons interacting with the
reggeons having fixed $j=1$. The gluon Regge trajectory
$j=j(t)$ appears
as an one-loop correction to the effective action for the bare reggeons.

It is convenient for us to use
the matrix representation of the Yang-Mills field
$v_\mu^a$:
\begin{eqnarray}
v_\mu\,\,=\,\,t^a\,v_\mu^a\,,
\end{eqnarray}
where the anti-hermitial matrices $t^a$ satisfy the commutation relations:
\begin{eqnarray}
[t^a,t^b]\,\,=\,\,f^{abc}\,t^c\,.
\end{eqnarray}
Here $f^{abc}$ are the structure constants and $T^a=it^a$ are the generators
of the colour group $SU(N_c)$ in the fundamental representation.

The virtual gluons in crossing channels, which lead to the Coulomb-like
interaction between the particles with a big difference in their rapidities
$y=\frac{1}{2}\ln (k^+/k^-)$ and which are reggeized after taking into
account
radiative corrections, are described by the fields
related  with the light-cone components of
the vector-potential $v_\mu$. Indeed, only the longitudinal part
$\sim \delta_{\mu \nu}^{\parallel}$ of the gluon
propagator gives a big contribution proportional to $s$. In accordance with
ref. [4] we denote these reggeon fields by $A_\pm$ . $A_\pm$  belong to the
adjoint representation of the group $SU(N_c)$ but they are considered
to be invariant under the gauge transformations for which the local
parameters $\chi (x)$ decrease at large $x$.

The particles which are
produced in direct channels can be arranged in the  groups consisting of
gluons within some rapidity intervals
$(y-\frac{\eta}{2},y+\frac{\eta}{2})$, where the auxiliary
parameter $\eta$ is chosen to be numerically big but significantly
smaller than the relative  rapidity of colliding particles $Y=\ln s$
\begin{eqnarray}
1\,\ll \,\eta \,\ll \,Y\,\,.
\end{eqnarray}

For the interactions among particles inside of each group
the introduced parameter $\eta$ is an
analog of the ultraviolet cut-off in the relative longitudinal momenta. For
the interactions between the neighbouring groups
$\eta$ plays the role of an infrared
cut-off. The $\eta$-dependence should disappear in the final result
analogously to the case of the normalization point dependence
in hard processes. In the leading logarithmic approximation all
transverse momenta $k_\perp$ of gluons in the Feynman diagrams are of the
same order of value as  transverse momenta $p_\perp$ of partons inside
of colliding hadrons [2]. This means that the c.m. pair energy $\sqrt s_i$
of the neighbouring gluons in the multi-Regge kinematics is significantly
bigger than $p_\perp$ and the effective parameter of the pertubation theory
is $g^2\ln (s/p_\perp^2)$. Beyond LLA
one should introduce the above parameter $\eta$
in the analogous inequality for the pair  energies of the neighbouring
clusters of particles
\begin{eqnarray*}
\ln \frac{s_i}{p_\perp^2}\,>\,\eta \,\,.
\end{eqnarray*}

We include this condition in the bare propagator of the reggeon
fields $A_\pm$ following the Gribov approach [12] for the pomeron case
\begin{eqnarray}
<A_+^{y'}(z^{\pm},\rho )A_-^y(0,0)>\,\,\sim \,\,\theta (y'-y-\eta)\,\,
\delta^2(z)\,\, \ln |\rho |\,.
\end{eqnarray}
Here $z^\pm $ and $\rho $ are correspondingly the  light-cone and transverse
components of the gluon coordinate $x_\mu$.
We took into account also that in the quasi-multi-Regge kinematics
the reggeon momenta
are transverse vectors ($q^2=q_\perp^2$) and in the effective vertices
one can consider
$A_+$ ($A_-$) as the fields independent of
$z_-$ ($z_+$) (cf.[4]):
\begin{eqnarray}
\partial_-A_+\,=\,\partial_+A_-\,=\,0\,.
\end{eqnarray}

The vector-potential $v$ can be presented in a simbolic form as
the following sum of its components  $V^y$ and $A_\pm^y$
describing correspondingly the gluons
in the direct and crossing channels with the rapidities $y$ inside of the
interval $\eta$:
\begin{eqnarray}
v\,\,=\,\,\sum\limits_{y}^{}(V^y\,+\,A^y).
\end{eqnarray}
Such representation is natural for the case, when $V$ and $A_\pm$ have the same
transformation properties.
However, according to our agreement $V$ and $A_\pm$ are transformed
differently under the gauge group.
Using the freedom of choosing an arbitrary
parametrization of $v$ in terms of $V$ and $A_\pm$
we shall modify later expansion (52) to satisfy the requirement of the
gauge invariance (see eq.(75)).

Because the interaction is local in the rapidity  except of the
long-range correlation
between $A_+$ and $A_-$ in (50),
we shall omit the index $y$ for all fields further on.
The usual Yang-Mills action for the gluons inside of each
rapidity interval ($y-\eta , y+\eta $) after using decomposition (52) takes
the form:

\begin{eqnarray*}
S_{YM}=-\int d^4x\,tr \{\frac{1}{2}G_{\mu \nu}^2 -[D_\mu ,G_{\mu -}]A_+ -
[D_\nu ,G_{\nu +}]A_-
+[D_\mu ,A_+][D_\mu ,A_-]-
\end{eqnarray*}
\begin{eqnarray*}
-\frac{1}{2}[D_-,A_-][D_+,A_+]+
\frac{g}{2}G_{+-}[A_-,A_+]-\frac{1}{4}[D_+,A_-]^2-
\frac{1}{4}[D_-,A_+]^2+
\end{eqnarray*}
\begin{eqnarray}
+\frac{g}{2}[D_+,A_-][A_-,A_+]+
\frac{g}{2}[D_-,A_+][A_+,A_-]-\frac{g^2}{4}[A_+,A_-]^2
\}\,\,,
\end{eqnarray}
where $D_\mu$ and $G_{\mu \nu}$ are respectively the  covariant
derivative and field strength tensor of the Yang-Mills field
\begin{eqnarray}
D_\mu =\partial _\mu +gV_\mu \,\,,\,\,
G_{\mu \nu}=\frac{1}{g}[D_\mu ,D_\nu]=
\partial_\mu V_\nu -\partial_\nu V_\mu +g[V_\mu ,V_\nu]\,.
\end{eqnarray}
The infinitesimal gauge transformations of $V$ and $G$ are
\begin{eqnarray}
\delta V_\mu = [D_\mu ,\chi ]\,\,,\,\,\,\delta G_{\mu \nu} =
g\,[G_{\mu \nu},\chi ]\,\,,
\end{eqnarray}
where $\chi$ is a small local parameter. Action (53) describes
various interactions of produced particles and reggeons. But we  know
from the
previous sections that one has to add to this action extra
terms to reproduce the induced vertices (28) and (45). They correspond to
the coherent emission of gluons belonging to a given rapidity interval
by neighbouring groups of particles and are needed in particular
to provide the gauge
invariance property of $A_\pm$ in accordance with the modified version (75)
of representation (52). We construct these induced terms in
the next sections.

\section{Effective action for quasi-elastic processes}

To begin with, let us consider in the effective action the linear terms
which describe quasi-elastic processes:
\begin{eqnarray}
S_1\,\,\,\,=-\int d^4 x \,\,tr\,\, [j_-\,\,A_+\,
+\,j_+\,\,A_-]\,.
\end{eqnarray}
Here each coefficient $j_\pm $ is the sum of  a modified Yang-Mills
current $j_\pm^{m\,YM}$  and an induced contribution $j_\pm^{ind}$:
\begin{eqnarray}
j_\pm =j_\pm^{m\,YM}+j_\pm^{ind}\,\,.
\end{eqnarray}
The Yang-Mills current $j_{\pm}^{YM}$
appearing in action (53)
is covariant under the gauge transformations (25)
\begin{eqnarray}
j_\pm^{YM}=-[\,D_\mu , G_{\mu \pm}\,]\,,\,\,\,\,\delta
j_\pm^{YM}\,=\,g\,[\,j_\pm^{YM},\chi \,]\,\,.
\end{eqnarray}
It  vanishes classically due to the equations of motion $\delta
G_{\mu \nu}^2=0$ being valid in the tree approximation. The
modified current $j^{m\,YM}$ is invariant under gauge transformations
(see (61)) and vanishes also on the mass shell.

The induced currents $j_-^{ind}$ and $j_+^{ind}$
depend only on $V_-$ and $V_+$,
respectively. Let us consider  for example $j_-^{ind}$ which describes the
gluon
production along $p_A$ . Using the reccurence relations (30)
being equivalent to
the requirement of  gauge invariance of the current $j^{ind}$, one can
construct $j_-^{ind}$ as a series in the QCD coupling constant

\begin{eqnarray}
j_-^{ind}(V_-) \,=\,\partial_{\perp \sigma}^2 \{V_- \,-
\,g\,V_- \frac{1}{\partial_-}V_- \,+\,
g^2\,V_- \frac {1}{\partial_-}V_- \frac{1}{\partial_-} V_-
\,-...\}=\partial_{\perp \sigma}^2\,\partial_-\,\frac{1}{D_-}\,V_- ,
\end{eqnarray}
where the transverse
Laplacian $\partial_{\perp \sigma}^2$ corresponds to the common factor
${\bf q}^2=-t$ in
the induced vertices (21, 27) for the quasi-elastic processes. The
infinitesimal gauge transformation of $j_-^{ind}$ is given by
\begin{eqnarray}
\delta j_-^{ind}=-\partial_{\perp \sigma}^2\{\partial_-\frac{1}{D_-}\chi
\partial_--\partial_-\chi \frac{1}{D_-}\partial_-\} =\partial_\sigma^2\,
\partial_-\,[\,\chi \,,\, \frac{1}{D_-}\,]\,\partial_-.
\end{eqnarray}
Using the
smallness of $\partial_\mp A_\pm $ (see (51)), we obtain from (60)
in accordance with reccurence
relations (30) , that effectively $\delta j^{ind}=0$ up to the terms
giving a vanishing contribution in expression (56) after integrating by
parts for $\chi$ which decreases as $x^\pm \rightarrow \infty$.
Therefore for
gauge invariance of $S_1$ (56) one should consider $A_\pm $ as gauge
invariant fields. In accordance with the invariant properties of
$j_\pm^{ind}$  one has to modify the Yang-Mills current (58) as follows
\begin{eqnarray}
j_\pm^{m\,YM}\,=\,U^{-1}(V_\pm )\,\,j_\pm^{YM}\,\,U(V_\pm )
\,\,,\,\,\,
U(V_\pm )=1-g\frac{1}{\partial_\pm }V_\pm
+g^2(\frac{1}{\partial_\pm }V_\pm )^2-...= \frac{1}{D_\pm }\partial_\pm \,\,.
\end{eqnarray}
Here $U(V_-)$ is the matrix transformed according to the fundamental
representation of the gauge group  $\delta U
=-g\chi U$, which provides $\delta j_\pm^{m\,YM}=0$. Such a modification is
possible, because both
$j^{YM}$ and $j^{m\,YM}$ are vanishing due to the equations of
motion and lead therefore  to the same production amplitude in the tree
approximation as discussed in section 2. In particular this yields the
cancellation of the poles
$1/\partial_\pm^k$ in $j_\pm^{m\,YM}$ which otherwise would
contradict the
requirement of absence of simultaneous singularities of the production
amplitudes in
the overlapping channels $s_i$ and $t$ . Note that this requirement is
fulfilled for $j^{ind}$ (59) due to the common factor
$\partial_{\perp \sigma}^2$ which cancels the pole $1/t$ .

Thus, under  the gauge transformation
(55) with $\chi$ decreasing as $x_\pm \rightarrow \infty $ the total current
(57) is invariant
\begin{eqnarray*}
\delta j_\pm =0\,\,.
\end{eqnarray*}
up to total derivatives $\partial_\pm$ giving due to (51) a negligible
contribution
to (56) for $\chi$ decreasing at $x^\pm \rightarrow \infty$. After
integrating by
parts in (56) the terms $\sim  g^0$ in eqs. (59) and (61) cancel and
the perturbative expansion for the total current $j$
starts with the contribution quadratic in $V$ and corresponding to effective
vertex (12)

\begin{eqnarray*}
j_- = g\{[V_\mu ,\partial_-V_\mu ]-[\partial_\mu V_\mu ,V_-]-
2[V_\mu , \partial_\mu V_-]-\partial_\mu ^2\,
V_-\frac{1}{\partial_-}V_-\,-
[\,j_-^{YM}\,,\,
\frac{1}{\partial_-}V_-\,]\}\,+
\end{eqnarray*}
\begin{eqnarray}
+g^2\,\{[V_\mu ,[V_-,V_\mu ]]+\partial_\mu^2 \,V_-\frac{1}{\partial_-}V_-
\frac{1}{\partial_-}V_-\,+\,[\,j_-^{YM}\,,\,\frac{1}{\partial_-}V_-\,]
\frac{1}{\partial_-}V_-\,\}\,
+\,O(g^3)\,,
\end{eqnarray}
where
\begin{eqnarray}
j_-^{YM}=\partial_\mu \partial_-V_\mu -\partial_\mu^2V_-+
g\,\{[V_\mu ,\partial_-V_\mu ]-[\partial_\mu V_\mu ,V_-]-
2[V_\mu ,\partial_\mu V_-]\}-g^2\,[V_\mu ,[V_\mu ,V_-]]
\end{eqnarray}
vanishes due to the equations of motion valid for quasi-elastic amplitudes in
the tree approximation.
Note that the quantity $A_\pm^y$  can be
considered as a classical component of the Yang-Mills field $v$.
We shall discuss this possibility later.

The singular coefficients $1/\partial_-^k$ in expansions (59) and (61)
are integral operators with unspecified boundary conditions.
A similar problem occurs in the Feynman diagram approach to  gauge
theories if one uses the light cone gauge $V_-=0$ because the gluon
propagator in this gauge is proportional to the tensor
\begin{eqnarray}
\Delta^{\mu \nu}= \delta^{\mu
\nu}- \frac{1}{\partial_-}(n_-^\mu \partial^\nu +n_-^\nu \partial^\mu ),
\end{eqnarray}
containing the pole at $\partial_-=0$. For this pole one can use the
Mandelstam-Leibbrandt prescription [13]
\begin{eqnarray}
\frac{1}{\partial_-}\rightarrow
\frac{\partial_+}{\partial_+\partial_--i\epsilon}.
\end{eqnarray}

In our case
due to the infrared cut-off (50) in the quasi-multi-Regge kinematics these
singularities are absent in the integration region. The appearence of
the poles in $\partial_-$ is related with the fact, that effective
action (56) is nonlocal because it is expressed in terms of the Wilson
contour integrals which depend generally on integration paths.

To clarify the last assertion we shall derive
the first term of action (56) using another method. To begin with, note
that in the
right light cone gauge $V_-^r=0$ where gluon propagators are proportional
to tensor (64) all induced interactions are not essential
and we can calculate the quasi-elastic amplitudes using
the first two terms of the usual Yang-Mills action (53). Furthermore, the
vector-potential $V^r$ in this gauge can be expressed in terms of the
vector-potential $V$ in an arbitrary gauge by the following gauge
transformation:
\begin{eqnarray}
V_\mu^r\,=\,U^{-1}(V_-)\,\,(V_\mu \,+\,\frac{1}{g}\partial_\mu
)\,\,U(V_-)\,\,,\, \,\,U(V_-)\,=\,P\,\exp
(-\frac{g}{2}\int\limits_{-\infty}^{x^-}d\,{x'}^-V_-),
\end{eqnarray}
where the operator $P$ implies the ordering of colour matrices in the
Wilson contour integral $U(V_-)$ in accordance with the increasing of the
field arguments  ${x'}^-$.

One can write down the operator representation for $U(V_-)$ and
$U^{-1}(V_-)$:
\begin{eqnarray}
U(V_-)\,=\,\frac{1}{1\,+\,g\,\partial_-^{-1}\,V_-}\,\,,\,\,U^{-1}(V_-)\,=
\,1\,+\,g\,\frac{1}{\partial_-}\,V_-\,\,,
\end{eqnarray}
where it is implied, that $U$ and $U^{-1}$ act on the unit constant matrix
from the left and right hand sides correspondingly.

The total current $j_-$ in the right gauge
contains only the term bilinear in fields $V^r$:
\begin{eqnarray}
j_-\,=\,j_-^{YM}(V^r)-\partial_\sigma \partial_-V_\sigma^r\,=
\,g\,[\,V_\sigma^r\,,\,\partial_-V_\sigma^r\,],
\end{eqnarray}
where $j_-^{YM}$ is given in eq.(58).

Inserting expression (66) for $V_\sigma^r$ in the
representation (68)  and using the gauge covariance
of $j^{YM}$ (58), we obtain  the total current in the gauge invariant form:
\begin{eqnarray}
j_-\,=\,U^{-1}(V_-)
\,j_-^{YM}\,U(V_-)\,-\,\partial_\sigma
\partial_-\,V_\sigma^r\,.
\end{eqnarray}
This result can be transformed to (57). Indeed, the first term is
$j^{m\,YM}$ (61) and the second term can be reduced to $j_-^{ind}$ (59)
using the smallness of the factor $\partial_-$ after integration by parts
in eq.(56) (see (51)) if it is not compensated by the
singularity $1/\partial_-$ in expression (66) for $V_{\sigma}^r$:
\begin{eqnarray}
-\partial_\sigma \,\partial_-\,
V_\sigma^r \rightarrow -\partial_{\perp \sigma} \,\partial_- \,
\frac{1}{g}\,\partial_{\perp \sigma}\,U(V_-)\,=\,j_-^{ind}(V_-)\,.
\end{eqnarray}

If the equations of motion for $V$ are fulfilled
the total current $j_\pm$
in (69) coincides with $j_\pm^{ind}$ (70) :
\begin{eqnarray}
j_\pm \rightarrow -\partial_\pm \partial_{\perp \sigma}^2\frac{1}{g}
U(V_\pm )\,.
\end{eqnarray}
We can use this expression for $j_\pm$ to construct the scattering amplitude
in the approximation of the quasi-elastic unitarity for the kinematics
where in the intermediate states of the $s$ and $u$ channel there are
two gluon clusters with fixed invariant masses moving in the
opposite directions along the momenta $p_A$ and $p_B$ . To begin with,
let us neglect all terms depending on $A_\pm$ in Lagrangian (53) except of
$tr (\partial_{\perp \sigma}A_+)(\partial_{\perp \sigma}A_-)$ and the
linear terms in (56) with substitution (71). After performing the
gaussian functional integration over $A_\pm$ leading to the gluon
propagators $1/\partial_{\perp \sigma}^2$ in the crossing channel, we obtain
the following effective action for this double quasi-elastic process:
\begin{eqnarray}
S_{doubl}=-\int d^4x\, tr \,\frac{1}{2}G_{\mu \nu }^2\,+\Delta S\,,
\,\,\Delta S\,=-\frac{1}{g^2}\int d^4x\,tr \,(\partial_-\partial_{\perp
\sigma}U(V_-))(\partial_+\partial_{\perp \sigma}U(V_+))    \,,
\end{eqnarray}
where $U(V_\pm )$ are the Wilson exponents (66).
The integrals over $x^\pm$ in the second term can be calculated and we
obtain for this contribution the
action of a two-dimensional $\sigma$-model
\begin{eqnarray}
\Delta S\,=\,-\frac{2}{g^2}\int d^2x_{\perp} \,\,tr \,\,(\partial_{\perp
\sigma}T(V_-))(\partial_{\perp \sigma}T(V_+))\,,
 \end{eqnarray}
where
\begin{eqnarray}
T(V_\pm )\,=\,P\,\,\exp (-\frac{g}{2}\,\int\limits_{-\infty}^{\infty} dx^\pm
V_\pm )\,.
\end{eqnarray}
This $\sigma$ model was earlier derived  by E. Verlinde and H. Verlinde [14]
using other arguments. Note,
that in our approach the Yang-Mills term $-\int d^4x\,tr \,\frac{1}{2}
G_{\mu \nu}^2$ responsible for the gluon
interactions incide of each of two produced clasters
is essential. Indeed, only due to the Euler-Lagrange equations
for the Yang-Mills action one can substitute $j$ by $j^{ind}$ in (57). Further,
even after solving the above $\sigma$ model we
should take into account non-eikonal corrections from the Yang-Mills
interactions in the intermediate states to obtain the
$S$-matrix with the full quasi-elastic unitarity.
For example, in QED the contribution analogous to action (73) is trivial:
\begin{eqnarray*}
\Delta S\,=\,\frac{1}{2}\,\int \,d^4x\,\,(\partial_{\perp \sigma}V_-)
(\partial_{\perp \sigma}V_+)
\end{eqnarray*}
and corresponds to the generalized eikonal approximation for scattering
amplitudes.
Nevertheless, it is known, that
in higher orders of the perturbation theory one should take into account
apart from the elastic
eikonal contribution also
the screening effects appearing due to the $e^+e^-$ pair production.
Action (72) can describe  quasi-elastic processes, but even in LLA
one should consider also the gluon emission in multi-Regge kinematics
(1) leading to
the gluon reggeization which can not be obtained in a simple way in
the framework of the above $\sigma$ model. In
the next sections we construct the other terms of the action which are
responsible for more general quasi-multi-Regge processes.

\section{Effective action for gluon production in the central rapidity
region}

As it was stressed above, there
is an ambiguity in expansion (52) of the total Yang-Mills field $v$ in
its components $V^y$ and $A^y$ which describe  particles in
direct channels and reggeized gluons in crossing channels, respectively.
Using in
fact this ambiguity we substituted the Yang-Mills current (58) by the
modified current (61) . This modified current $j^{m\,YM}$ appears in
the action
as a coefficient in front of the linear term in the expansion over $A_\pm$
if we chose instead of (52) the different decompositions for each
Lorentz projection of $v_\mu$
\begin{eqnarray}
v_{\perp \mu}=\sum\limits_{y}^{}\,V_{\perp \mu}\,,\,\,\,\,
v_\pm = \sum\limits_{y}^{}\,\{\,V_\pm ^y\,+\, U(V_\mp )\,A_\pm
^y\,U^{-1}(V_\mp )\,\}\,, \end{eqnarray}
where $U(A_\pm)$ are determined in eqs (61, 66) .

In particular, using representation (75) and results (37), (42) and (45) from
our discussion of the gluon production in the central rapidity region
we can write the corresponding action bilinear in fields $A_\pm$ in the
following form  (cf. (56) and (57))
\begin{eqnarray}
S_2\,\,=\,-\int d^4x\,\,(\,L_2^{m\,YM}\,+\,L_2^{m\,ind}\,)\,\,.
\end{eqnarray}

The Lagrangians $L^{m\,YM}$ and $L^{m\,ind}$ are certain modifications of
the Yang-Mills contribution (see (53))
\begin{eqnarray}
L_2^{YM}\,=\,tr \{\,[\,D_\mu , A_+\,]\,[\,D_\mu ,
A_-\,]\,-\frac{1}{2}\,[\,D_- ,
A_- \,]\,[\,D_+ , A_+\,]\,+\,\frac{g}{2}\,G_{+-}\,[\,A_- , A_+\,]\,\}\,\,
\end{eqnarray}
and of the induced contribution (cf.(45)) related to the
induced currents $j_\pm^{ind}$ (59) as
\begin{eqnarray*}
L_2^{ind}\,=\,tr \,A_-\,\frac{\partial}{\partial \,V_-}\,tr \,
\{j_-^{ind}(V_-)-\partial_{\perp \sigma}^2V_-\}\,A_+\,+
\end{eqnarray*}
\begin{eqnarray}
+\,tr \,A_+\,\frac{\partial}{\partial \,V_+}\,tr \,
\{j_+^{ind}(V_+)-\partial_{\perp \sigma}^2V_+\}\,A_-\,\,,
\end{eqnarray}
where it is implied, that after differentiating $j_-$ and $j_+$ over
$V_-$ and $V_+$, respectively, the fields $A_-$ and $A_+$ substitute the
fields $V_-$ and $V_+$ at the corresponding empty positions. In (78) we
subtracted
the kinetic terms because they appeared already in $L_2^{YM}$ as
$\,tr \,(\partial_{\perp \mu}A_+)
(\partial_{\perp \mu}A_-)\,$. The
modification of contributions (77, 78)
is needed to provide the gauge invariance of action (76) in
accordance with our prescription for the fields $A_\pm $ as invariants of
gauge transformations (55) with $\chi$ decreasing as $x^\pm \rightarrow
\infty$ . In the case of the YM term (77) this invariance leads to the
necessity of the redefinition of  the reggeon fields $A_\pm$ in accordance
with new decomposition (75) of fields $v$
\begin{eqnarray}
A_\pm \,\,\rightarrow \,\,U(V_\mp )\,A_\pm \,U^{-1}\,(V_\mp )\,\,.
\end{eqnarray}
This procedure gives the following
result for the modified Yang-Mills contribution $L^{m\,YM}$ after using
the covariance properties of $D_\mu$ and eq.(51) for the
derivatives $\partial_\mp A_\pm$ :

\begin{eqnarray*}
L_2^{m\,YM}=tr \,\{\,[\,\tilde{D}_{\perp \mu}^{(-)}\,,\,
A_+\,]\,\,W\,\,[\,\tilde{D}_{\perp \mu}^{(+)}\, , \,A_-\,]\,\frac{1}{W}\,+
\end{eqnarray*}
\begin{eqnarray}
+\,g\,\,tr \,
\{\,(\partial_+\tilde{V}_-^{(+)})\,A_-\,\frac{1}{W}\,A_+\,W\,+\,
(\partial_-\tilde{V}_+^{(-)})\,A_+\,W\,A_-\,\frac{1}{W}\}.
\end{eqnarray}
Here the gauge invariant matrix $W$ is given by
\begin{eqnarray}
W\,=\,U^{-1}(V_-)\,U(V_+)\,=
\,(1+\frac{g}{\partial_-}V_-)(1+\frac{g}{\partial_+}V_+)^{-1}\,
\end{eqnarray}
and the quantities $\tilde{D}^{(\pm)} , \tilde{V}^{(\pm)}$ are determined
as
\begin{eqnarray}
\tilde{D}_\mu^{(\pm)}=U^{-1}(V_\pm)\,D_\mu
\,U(V_\pm)\,,\,\,\,\tilde{V}_\mu^{(\pm)}\,= \,U^{-1}(V_\pm)\,V_\mu
\,U(V_\pm)\,\,. \end{eqnarray}

Now let us consider $L_2^{ind}$ (78). Its modification which is compatible
with the gauge invariant properties of $S_2$  is given by
\begin{eqnarray*}
L_2^{m\,ind}=tr \,U(V_+)A_-U^{-1}(V_+)\frac{\partial}{\partial V_-}
tr \,j_-^{ind}A_++
\end{eqnarray*}
\begin{eqnarray}
+tr \,U(V_-)A_+U^{-1}(V_-) \frac{\partial}{\partial V_+} tr \,j_+^{ind}
(V_+)A_-
-tr\,\{A_-\partial_{\perp \sigma}^2A_+\,+\,A_+\partial_{\perp \sigma}^2A_-\}.
\end{eqnarray}
This representation is again in accordance  with our redefinition (79) of
the fields $A_\pm$ in action (53) including now
apart from the Yang Mills contributions also the induced
terms. $L^{m\,ind}$ (83) is constructed in
a such way that it can be obtained from the linear terms in (56)
if we substitute the fields $V_\pm$ in eq.(59) for $j_\pm^{ind}$ by the
total
field $v_\pm$ and expand the result up to the linear term in $A_\pm$ after
using the new decomposition (75) of fields $v$

\begin{eqnarray}
j_\pm^{ind}(V_\pm) \rightarrow
j_\pm^{ind}(v_\pm)-j_\pm^{ind}(A_\pm)\,=\,
j_\pm^{ind}(V_\pm )+U(V_\mp )A_\pm U^{-1}(V_\mp)
\frac{\partial}{\partial \,V_\pm}j_\pm ^{ind}(V_\pm)+...\,\,.
\end{eqnarray}

The perturbative expansion of the Lagrangian in action (76) into a series
over $V_\mu$ is given by
\begin{eqnarray*}
L_2^{m\,YM}+L_2^{m\,ind}=tr \,[(\partial_{\perp \mu}A_+)(\partial_{\perp
\mu}A_-)+
\end{eqnarray*}
\begin{eqnarray*}
+g\,\{-V_{\perp \mu}\,[\,\partial_{\perp
\mu}A_+\,,\,A_-\,]\,+\,V_{\perp \mu}\,[\,A_+\,,\,\partial_{\perp
\mu}A_-\,]\,
+\frac{1}{2}(\partial_-V_+-\partial_+V_-)[A_+,A_-]-
\end{eqnarray*}
\begin{eqnarray*}
-[\partial_\mu^2A_+,A_-]\frac{1}{\partial_-}V_-+
[A_+,\partial_\mu^2A_-]\frac{1}{\partial_+}V_++K_1\,\}\,+
\end{eqnarray*}
\begin{eqnarray*}
+g^2\,\{\,[V_{\perp \mu},A_+][V_{\perp \mu},A_-]+
\frac{1}{2}[V_-,A_+][V_+,A_-]-
\frac{1}{2}[V_+,V_-]\,[A_+,A_-]+
\end{eqnarray*}
\begin{eqnarray*}
+(\partial_{\perp \mu}^2A_+)\left(A_-\frac{1}{\partial_-}
V_-\frac{1}{\partial_-}V_-+V_-\frac{1}{\partial_-}A_-\frac{1}{\partial_-}V_-+
V_-\frac{1}{\partial_-}V_-\frac{1}{\partial_-}A_-\right)+
\end{eqnarray*}
\begin{eqnarray}
+(\partial_{\perp \mu}^2A_-)\left(A_+\frac{1}{\partial_+}
V_+\frac{1}{\partial_+}V_++V_+\frac{1}{\partial_+}A_+\frac{1}{\partial_+}
V_++V_+\frac{1}{\partial_+}V_+\frac{1}{\partial_+}A_+\right)\,+\,K_2\,\}\,
+\,O(g^3)\,]\,.
\end{eqnarray}
The total contribution of the sum
$K=\sum\limits_{r=1}^{\infty}g^rK_r$
appearing in (85)  and of the  terms in (62) which contain
$j_\pm^{YM}$
can be written as a perturbative expansion  of the following expression:
\begin{eqnarray*}
\Delta L_2\,=\,L_2^{YM}(V+UAU^{-1})-L_2^{YM}(V+A)+
\end{eqnarray*}
\begin{eqnarray}
tr \,\{A_-[j_+^{ind}(V+UAU^{-1})-j_+^{ind}
(V+A)]+A_+[j_-^{ind}(V+UAU^{-1})-j_-^{ind}(V+A)]\}.
\end{eqnarray}
Here we have used short-hand notations for two different representations (52)
and (75) of the total field $v$.
Expression (86) is zero due to the equations of motion for the field $V$
because it is a difference of the same effective Lagrangian for two various
parametrizations of $v$.
In the next section we shall discuss  the corresponding effective action
$S_{eff}(v)$ for quasi-multi-Regge
processes in QCD for an arbitrary parametrization.

\section{Reggeon calculus in QCD }

{}From the results of the previous sections we conclude that the
gauge-invariant
action for the gluon-reggeon interactions which are local in each
rapidity interval ($y-\eta ,y+\eta $)  can be written in terms
of the Yang-Mills field $v$ and the reggeon fields $A_\pm$ as follows
\begin{eqnarray}
S_{eff}(v,A_\pm )=-\int d^4 x \,\,
tr \,\{\frac{1}{2}G_{\mu \nu}^2(v)\,+\,[A_-(v_-)-A_-]j_+^{reg}\,+\,
[A_+(v_+)-A_+]j_-^{reg}\}\,.
\end{eqnarray}
Here $A_\pm (v_\pm)=(1/{\partial_{\perp \sigma}^2})j_\pm^{ind}(v_\pm)$
are composite fields (see (70)):
\begin{eqnarray*}
A_\pm (v_\pm)\,=\,v_\pm -gv_\pm \frac{1}{\partial_\pm}v_\pm +
g^2v_\pm \frac{1}{\partial_\pm}v_\pm \frac{1}{\partial_\pm}v_\pm -...\,=\,
-\frac{1}{g}\partial_\pm U(v_\pm)
\end{eqnarray*}
and $j_\pm^{reg}=j_\pm ^{reg}(A_\pm)$ are
reggeon currents satisfying the kinematical constraints
$\partial_\mp j_\pm^{reg}=0$ (see (51))
which are important for the gauge invariance of
action (87). Note, that expression (87) is similar to the first Legendre
transformation from the Yang-Mills action,
but instead of its stationarity  in $j_\pm ^{reg}$
one should use
the on-shell condition for $A_\pm$:
\begin{eqnarray*}
j_\pm^{reg}\,=\,\partial_{\perp \sigma}^2\,A_\pm \,
\end{eqnarray*}
to express $j_\pm ^{reg}$ through $A_\pm$. It is related with the fact,
that action (87) describes only interactions in the given rapidity interval
and the interactions between the particles with different rapidities will be
taken into account in the framework of the reggeon field theory.

One can add to action (87)
also the quark contribution
$\int d^4x \, \bar{\psi}(i\hat{D}-m)\psi$. It lead to an additional term
$g\bar{\psi}\gamma_\sigma \psi$ in the Yang-Mills current
appearing in classical equations (88) without other
essential modifications of the theory. The more complicated effective action
including also the fields $a_\pm$ describing the reggeized quarks
can be constructed for the backward
scattering processes with the
fermionic exchange in the crossing channel (cf.[5]). Note that for
the case of the electroweak theory one should add to action (87)
the terms which are responsible
for interactions of quarks, leptons and the Higgs
particle (cf.[2]).

Now we want to consider the problem of constructing the Reggeon calculus in
QCD starting from the effective action (87).
Because $A_\pm (v_\pm )$ has a linear term in $v_\pm$  the
classical extremum of $S_{eff}$ (87) is situated at non-vanishing values of
$v=\tilde{v}$ satisfying the gauge-invariant Euler-Lagrange equations
\begin{eqnarray}
j_{\perp \sigma}^{YM}(v)=0\,\,,\,\,j_\pm^{YM}(v)=-\frac{\partial}{\partial
v_\mp} tr \,A_\mp (v_\mp)\,j_\pm ^{reg} = -U(v_\mp)\,(\partial_{\perp
\sigma}^2A_\pm) \,U^T(v_\mp)\,\,.
\end{eqnarray}
Here the matrix $U(v_\mp)$ is determined in eqs.(66, 67) and the
transposed matrix $U^T(v_\mp)$ is
\begin{eqnarray*}
U^T(v_\mp)\,=\,\frac{1}{1+g\,v_\mp \partial_\mp ^{-1}}\,
\end{eqnarray*}
which is supposed to be multiplied by a constant unit matrix from the left
hand side.

Due to  the invariance properties of $A_\pm$ under
the solution $\tilde{v}$ of eqs.(88) is
degenerate.
We use here the Feynman gauge in which the gluon propagator of the field
$v$ is $\delta^{\mu \nu}/k^2$. Taking into account
that $\partial_+A_-$ and $\partial_-A_+$ are negligible in the multi-Regge
kinematics (see (51)) one can construct the following perturbative
solution of (88)
\begin{eqnarray*}
\tilde{v}_+\,=\,A_+\,+\,g\,\partial_\mu^{-2}\,\{\,\,[(\partial_{\perp
\sigma}^2A_+)\,,\,(\partial_-^{-1}A_-)]\,-\,
\frac{1}{2}[A_-,\partial_+A_+]\,\}\,+\,O(g^2)\,\,,
\end{eqnarray*}
\begin{eqnarray*}
\tilde{v}_-\,=\,A_-\,+\,g\,\partial_\mu^{-2}\,\{\,[(\partial_{\perp
\sigma }^2A_-)\,,\,(\partial_+^{-1}A_+)]\,
-\,\frac{1}{2}[A_+,\partial_-A_-]\,\}\,+\,O(g^2)\,\,,
\end{eqnarray*}
\begin{eqnarray}
\tilde{v}_{\perp \sigma}\,=\,\frac{1}{2}\,\,g\,\,
\partial_\mu^{-2}\,\{[A_+,\partial_{\perp
\sigma}A_-]+[A_-,\partial_{\perp \sigma}A_+]
\}\,+\,O(g^2)\,\,.
\end{eqnarray}

The effective  action calculated at the classical solution
$v=\tilde{v}$
\begin{eqnarray*}
\tilde{S}_{eff}(A_\pm)=-\int d^4x \,\,tr \,\{\frac{1}{2}(\partial_{\perp
\sigma}A_-)(\partial_{\perp \sigma }A_+)\,-\,
\end{eqnarray*}
\begin{eqnarray}
-\frac{1}{2}\,g\,\left(\,
(\partial_{\perp \sigma}^2A_-)\,[(\partial_+^{-1}A_+)\,,\,A_+]\,+\,
(\partial_{\perp \sigma}^2A_+)\,[(\partial_-^{-1}A_-)\,,\,A_-]\,
\right)\,+\,O(g^2)\}
\end{eqnarray}
describes generally all possible self-interactions of the Reggeon fields
$A_\pm$ in the
tree approximation. In particular, the tri-linear term is responsible for
the transition of the reggeon corresponding to the field $A_\pm$ into
the state constructed from two reggeons described by fields
$A_\mp$.
This transition is suppressed for the case of the elastic amplitude
according to the Gribov signature conservation
rule, because the signature of the reggeized gluon is negative.
Nevertheless, it was argued [15,16], that the inclusion of this triple
reggeon vertex simplifies the reggeon calculus and clarifies the mechanism
of the gluon reggeization.
Note that this vertex is proportional to $q^2$,
and contains the singularity $\partial_\pm^{-1}$ corresponding to the
contribution of diagrams in which
there are highly virtual particles in the direct channels [5]. The
quadri-linear
term contains the transition of one reggeon into the state constructed from
three reggeons [16]. This transition is not suppressed by the Gribov
signature conservation rule.
There is also a contribution which describes the scattering of
two reggeons and gives in particular  the
integral kernal for the BFKL equation [2]. The six-linear term leads to
to the pomeron self-interactions which are responsible for screening
corrections [16]. All these terms can be obtained using the perturbative
solution (89) of classical equations  for the effective action (87). We hope
to return to these problems in future publications.

To calculate higher loop corrections to the Reggeon Lagrangian
one should write the field $v$ as a sum of the
classical field $\tilde{v}$ and the field $\epsilon$ describing quantum
fluctuations near the classical field (cf.(52, 75)):
\begin{eqnarray}
v\,=\,\tilde{v}\,+\,\epsilon \,\,,
\end{eqnarray}
expand the action in $\epsilon $ (cf(53)):
\begin{eqnarray*}
\Delta S=S_{eff}-\tilde{S}_{eff}\,=\,-\int d^4 x \,tr \,\{\,[D_\mu
,\epsilon_\nu]^2\,-\,[D_\mu ,
\epsilon_\nu ][D_\nu , \epsilon_\mu ]\,+\,g\,G_{\mu \nu}\,[\epsilon_\mu ,
\epsilon_\nu ]+
\end{eqnarray*}
\begin{eqnarray}
+\frac{1}{2}(\epsilon_-\frac{\partial}{\partial v_-})(\epsilon_-
\frac{\partial}{\partial v_-})j_-^{ind}(v_-)A_++
\frac{1}{2}(\epsilon_+\frac{\partial}{\partial
v_+})(\epsilon_+\frac{\partial}{\partial v_+})j_+^{ind}(v_+)A_-
+O(\epsilon^3)\,\}.
\end{eqnarray}
and calculate the functional integral over the quantum fluctuations
$\epsilon$. Using the gaussian approximation one can find in particular
the one-loop correction to the BFKL kernal in an independent way in
comparison with the dispersion method of refs.[3],[11]. The possible
advantage of this approach is a better infrared convergency of
intermediate expressions. To calculate the two-loop correction to
the gluon Regge trajectory one should expand $S_{eff}$ up to $\epsilon^4$
taking into account the terms bilinear in $A_\pm$.
These important problems will be discussed elsewhere . Here we
use expressions (89)  to find $\Delta S$ (92)
only up to quadratic terms in $\epsilon$ and bilinear in $A_\pm$:
\begin{eqnarray*}
\Delta S=-\int d^4x tr \{(\partial_\mu \epsilon_\nu
)^2-(\partial_\mu \epsilon_\nu )(\partial_\nu \epsilon_\mu )
+g\{2(\partial_\mu \epsilon_\nu )[\tilde{v}_\mu ,\epsilon_\nu ]-
2(\partial_\nu \epsilon_\mu )[\tilde{v}_\mu ,\epsilon_\nu ]+
\end{eqnarray*}
\begin{eqnarray*}
+2(\partial_\nu
\tilde{v}_\mu )[\epsilon_\nu ,\epsilon_\mu ]
-(\partial_{\perp
\sigma}^2A_+)\epsilon_-\frac{1}{\partial_-}\epsilon_--(\partial_{\perp
\sigma}^2A_-)\epsilon_+\frac{1}{\partial_+}\epsilon_+\,\}\,+\,
\end{eqnarray*}
\begin{eqnarray*}
+g^2\{[A_+,\epsilon_\nu][A_-,\epsilon_\nu]-\frac{1}{2}[A_+,\epsilon_+]
[A_-,\epsilon_-]-\frac{1}{4}
[A_+,\epsilon_-]^2-\frac{1}{4}[A_-,\epsilon_+]^2-
\end{eqnarray*}
\begin{eqnarray*}
-\frac{1}{2}[A_+,A_-][\epsilon_+,\epsilon_-]
+(\partial_{\perp
\sigma}^2A_+)(\epsilon_-\frac{1}{\partial_-}\epsilon_-\frac{1}{\partial_-}
A_-+\epsilon_-\frac{1}{\partial_-}A_-\frac{1}{\partial_-}\epsilon_-+
A_-\frac{1}{\partial_-}\epsilon_-\frac{1}{\partial_-}\epsilon_-)\,+
\end{eqnarray*}
\begin{eqnarray}
+(\partial_{\perp
\sigma}^2A_-)(\epsilon_+\frac{1}{\partial_+}\epsilon_+\frac{1}{\partial_+}A_++
\epsilon_+\frac{1}{\partial_+}A_+\frac{1}{\partial_+}\epsilon_++
A_+\frac{1}{\partial_+}\epsilon_+\frac{1}{\partial_+}\epsilon_+)\}\}\,,
\end{eqnarray}
 where one should substitute $\tilde{v}$ by expressions (89). The terms
bilinear simultaneously in $A_\pm$ and in $\epsilon$
can be used for finding next to leading corrections to the BFKL
pomeron.  We consider here only the contributions
which are linear in $A_\pm$ and contain the
singularities $\partial_\pm^{-1}$. Expanding
the integrand $exp \,(-iS_{eff})$ in the
functional integral over $\epsilon$
up to the higher order terms containing linearly both $A_+$ and $A_-$
and substituting the products of $\epsilon_+$ and $\epsilon_-$ by
the free propagators in the Feynman gauge
\begin{eqnarray}
<\epsilon_-(x) \epsilon_+(0)>\,=\,-2i\int
\frac{d^4k}{(2\pi)^4}\,\frac{1}{k^2+i\,0}\,exp(ikx)\,\, ,
\end{eqnarray}
we obtain the corresponding one-loop contribution to the effective action
in the form (cf.(51))
\begin{eqnarray*}
S_{eff}^1=-\int d^4x \, \int d^2x'_\perp
\,\int d\,y \,
tr\,(-\frac{1}{2})(\partial_{\perp \sigma}A_-^y(x_-,x_\perp))\,\,
(\partial_{\perp
\sigma}A_+^{y}(x_+,x'_\perp))\,
\end{eqnarray*}
\begin{eqnarray}
\frac{g^2}{(2\pi)^3}\,N_c\,
\{\frac{\theta(|x_\perp-x'_\perp|-\delta)}{|x_\perp-x'_\perp|^2}\,-
\,2\,\pi \delta^2(x_\perp - x'_\perp)\, ln \frac{1}{\delta \lambda}\}\,,
\end{eqnarray}
where $\delta \rightarrow 0$ and $\lambda$ is proportional to a fictious
gluon mass, introduced to remove the infrared divergency. The fields
$A_\pm^y$ are determined in section 4 (see (52)). The appearence
of the integral over the rapidity $y=\frac{1}{2}\ln (k^+/k^-)$ is related
with the fact, that due to the singularities $\partial_\pm^{-1}$
corresponding to the propagators of intermediate gluons in the direct
channel we calculate really the contribution of the box diagram which
contains $\ln s$.
The Fourie transform of the expression in the last line of (95)
gives just the gluon Regge trajectory (3)
\begin{eqnarray}
\omega (t)=\frac{g^2}{(2\pi)^3}\,N_c\,\int d^2x_\perp \exp (iqx_\perp)\,
\{\frac{\theta
(|x_\perp|-\delta)}{|x_\perp|^2}\,
-\,2\,\pi\,\delta^2(x_\perp)\,\ln\frac{1}{\delta \lambda}\}\,.
\end{eqnarray}
In accordance with our agreement (50) concerning the
regularised propagator of the
reggeon fields $A_\pm$ we modify  their kinetic term in expression
(90)
\begin{eqnarray}
S^{kin}(A_\pm)=-\int d\,y \int d^4x\,tr\, \frac{1}{2}\,(\partial_{\perp
\sigma}A_-^y) \,\frac{\partial}{\partial \,y}\,
(\partial_{\perp \sigma}A_+^{y+\eta}).
\end{eqnarray}
It leads to an additional factor $\theta (y'-y-\eta)$
in the free correlation function of the fields $A_+^{y'}$ and $A_-^y$.
The dependence on the auxiliary parameter $\eta$ should disappear in the
final result, as it was argued in section 4. With taking into account
the one loop correction $S_{eff}^1$ the renormalized correlation function
corresponding to the sum of expressions (95) and (97) contains the
Regge factor $\exp (\omega(q^2)\,(y'-y))$ in the momentum representation:
\begin{eqnarray}
\int d^2\,x_\perp \,\exp (ix_\perp q)\,<A_+^{y'}(x_\perp)A_-^y(0)>_{ren}
\sim \theta(y'-y-\eta)\,\exp (\omega(q^2)\,(y'-y))\,\,.
\end{eqnarray}
It is well known [2], that the infrared divergency in the gluon Regge
trajectory (96) at $\lambda \rightarrow 0$ is cancelled in the BFKL
kernel with the contribution corresponding to the real gluon emission.
This contribution appears in the classical reggeon action
(90) as a quadri-linear term in $A_\pm$. We shall discuss the mechanism
of this cancellation in the framework of this
functional approach somewhere
else.

Note, that if one wants to construct not only the reggeon
vertices, but
to compute also the production amplitudes in the quasi-multi-Regge
kinematics in the tree approximation (see sections 5 and 6), the effective
field theory describing all possible interactions of the reggeon fields
$A_\pm$ and the particle fields $V_\mu$ can be derived from the action
(87) using the known functional methods for calculating the $S$-matrix
elements [18]. For this purpose one should find the
solution of classical equations (88) with a fixed asymptotic behaviour
$V_\mu$ for $v_\mu$ at $t \rightarrow \pm \infty$ and put this solution in
expression (87), obtaining the generating functional for the $S$-matrix
elements. Futhermore, the reggeon-reggeon-particle vertex in
one loop approximation [3] can be calculated by finding the contribution
from the quantum fluctuations near the classical solutions.

\section{Conclusion}

In this paper we constructed the gauge-invariant effective action (87)
describing the interaction of  bare reggeized gluons and
physical gluons within the rapidiy interval $\eta$.
By eliminating the Yang-Mills field $v_\mu$ with the
use of equations of motion (88) the reggeon action
(90) is derived in the tree approximation of
perturbation theory. One can calculate one loop corrections to this action,
which leads in particular to the gluon reggeization (98). One
loop corrections to the BFKL kernel are urgently needed for the
consistent theoretical description of the small-x structure functions
measured at HERA.

The possibility to represent the initial Yang-Mills
theory in the form of a reggeon field model with the effective vertices
calculated perturbatively
is important, because the $s$-channel unitarity of the $S$-matrix for
the theory with action (87) is
transformed into various relations among the reggeon vertices. Therefore
the multi-reggeon dynamics in the crossing channel turns out to be in the
agreement with the unitarity constraints in the direct channels. The reggeon
calculus can be presented as a field theory in the 3-dimensional space [12]
where the two-dimensional coordinates $\rho =x_\perp$ describe the impact
parameters of the reggeons and the time coordinate $y$ is their rapidity.
The Schr\"odinger equation for the wave function
$\Psi_\omega^n(\rho_1,\rho_2,...\rho_n)$ of the
colourless glueball with a complex spin $j=1+\omega$ includes generally
the components with an arbitrary number $n$ of the reggeized gluons and
the transitions  between these states
can be obtained from action (87).
However, in the generalised
leading logarithmic approximation the number of the reggeised gluons
is conserved. The BKP
equations [6], obtained in this approximation, have a number of
remarkable properties in the multi-colour QCD [7-10]. One can
believe that at least some of these properties will survive
in the above reggeon field theory.

The method of constructing the effective action, describing the
quasi-multi-Regge processes in the Yang-Mills theory can be generalised
to the case of quantum gravity, where the tri-linear effective vertices
for the reggeon-particles interactions are known [18]. For the multi-Regge
kinematics the action was given earlier [4]. It was used for finding
the scattering amplitudes at super-Planck energies [19]. Recently this
action was derived from the Hilbert-Einstein action by integrating over
the heavy modes [20]. The action responsible for the quasi-multi-Regge
processes in gravity can be written in the form similar to eq.(87):
\begin{eqnarray}
S=I_G+I_{ind}=-\frac{1}{16\pi G}\int
d^4x\,\sqrt{g(x)}\,[R\,-\frac{\sqrt{8\pi
G}}{4}(R_{++}^{ind}A_{--}+R_{--}^{ind}A_{++})]\, \end{eqnarray}
where $I_G$ is the Hilbert-Einstein action, the fields $A^{++}$ and $A^{--}$
describe the  regeized gravitons in the $t$-channel. The quantities
$R_{++}^{ind}$ and $R_{--}^{ind}$ are the light-cone components of the Richi
tensor $R_{\mu \nu}$ calculated in  the linear approximation over the
graviton fields
$h^{\rho \sigma}=g^{\rho \sigma}-\delta^{\rho \sigma}$
in the left ($h^{-\sigma}=0$) and right ($h^{+\sigma}=0$) gauges
correspondingly. The discussion of this action will be given somewhere
else.

\vspace{1cm}
\noindent
{\large \bf Acknowledgements}\\
I would like to thank the Alexander von Humboldt Foundation for the award,
which gave me the possibility to work on this problem at
DESY-Hamburg and DESY-Zeuthen. The fruitful discussions with Ya. Balitsky,
J. Bartels, J. Bl\"umlein, R. Kirschner and L. Szymanowski were very
helpful.

\newpage
\noindent
{\Large \bf Appendix}\\
Below we write down the results of ref. [11] for production amplitudes of
quasi-multi-Regge processes in the Born approximation using slightly
different notations and
calculate corresponding matrix elements in the light-cone
gauge.  To begin with, let us consider the quasi-elastic process of one
gluon emission in the fragmentation region of the colliding gluon $A$. The
momenta of the final state gluons are $k_1$,$k_2$ and $p_{B'}$, the
momenta of two initial
gluons are $-k_0$ and $p\equiv p_B$. The corresponding Lorentz and colour
indices are $a_1,a_2,a_0,B',B$ and $\nu_1,\nu_2,\nu_0,\beta',\beta$. We
introduce the invariants:
\begin{eqnarray*}
t=(k_0+k_1+k_2)^2 \equiv q^2\,,\, s_{12}=(k_1+k_2)^2\,,\,
t_1=(k_0+k_1)^2\,,\,
t_2=(k_0+k_2)^2 \end{eqnarray*}
and the Sudakov components of momenta $k_1,k_2$:
\begin{eqnarray*}
\beta_{1,2}=\frac{(k_{1,2}\,p)}{(p_A\,p)}\,,\,\, \beta_0=-1\,\,.
\end{eqnarray*}
Then the production amplitude given by eq. (22) from the second paper in
ref.[11] can be written in the form of  eq. (17) with
\begin{eqnarray*}
\phi_{a_0a_1a_2c}^{\nu_0\nu_1\nu_2+}\,=\,
8t\sqrt{s}\{T_{a_1a_0}^{d}T_{a_2d}^{c}a^{\nu_0\nu_1\nu_2}\,+\,
\left([a_1,\nu_1, k_1] \leftrightarrow [a_2,\nu_2, k_2] \right)\}\,\,,
\end{eqnarray*}
where according to eq. (23) from ref.[11] the tensor $a^{\nu_0\nu_1\nu_2}$
is given by
\begin{eqnarray*}
a^{\nu_0\nu_1\nu_2}=
\end{eqnarray*}
\begin{eqnarray*}
=\frac{p^{\nu_0}}{s}
\left(\frac{p^{\nu_1}p^{\nu_2}}{\beta_2s^2}-
\frac{p^{\nu_1}k_1^{\nu_2}-k_2^{\nu_1}p^{\nu_2}}{s_{12}\,s}-
\frac{k_0^{\nu_1}p^{\nu_2}}{\beta_2\,s\,t_1}+\frac{k_0^{\nu_1}k_1^{\nu_2}}{t}
(\frac{1}{t_1}+\frac{1}{s_{12}})
-\frac{k_2^{\nu_1}k_0^{\nu_2}}{s_{12}\,t}+
\frac{k_0^{\nu_1}k_0^{\nu_2}}{t\,t_1}\right)+
\end{eqnarray*}
\begin{eqnarray*}
+k_1^{\nu_0}\left(\frac{p^{\nu_1}p^{\nu_2}}{\beta_2s^2t_1}+
\frac{(k_2^{\nu_1}p^{\nu_2}-p^{\nu_1}k_1^{\nu_2})}{s\,t}(\frac{1}{t_1}+
\frac{1}{s_{12}})-\frac{p^{\nu_1}k_0^{\nu_2}}{s\,t_1\,t}\right)-
\end{eqnarray*}
\begin{eqnarray*}
-k_2^{\nu_0}
\left(\frac{p^{\nu_1}k_1^{\nu_2}-k_2^{\nu_1}p^{\nu_2}}{s_{12}\,s\,t}+
\frac{k_0^{\nu_1}p^{\nu_2}}{s\,t_1\,t}\right)-
\end{eqnarray*}
\begin{eqnarray*}
-\frac{\delta^{\nu_1\nu_2}}{2}
\left(\frac{(t\beta_2-t_2)p^{\nu_0}}{s_{12}\,s\,t}+
\frac{\beta_2}{t}(\frac{1}{t_1}+
\frac{1}{s_{12}})k_1^{\nu_0}-
\frac{\beta_1k_2^{\nu_0}}{s_{12}\,t}\right)-
\end{eqnarray*}
\begin{eqnarray*}
-\frac{\delta^
{\nu_0\nu_1}}{2}\left((\frac{t}{t_1\,\beta_2}-
\frac{t_2}{t_1})\,\frac{p^{\nu_2}}{s\,t}-
\frac{\beta_1}{t_1t}\,k_0^{\nu_2}-(\frac{1}{t_1}+
\frac{1}{s_{12}})\frac{k_1^{\nu_2}}{t}\right)-
\end{eqnarray*}
\begin{eqnarray*}
-\frac{\delta^{\nu_0\nu_2}}{2}\left(-\frac{p^{\nu_1}}{st}-
\frac{\beta_2k_0^{\nu_1}}{t_1t}+
\frac{k_2^{\nu_1}}{s_{12}t}\right)\,\,.
\end{eqnarray*}
One can verify that this expression for
$\phi_{a_0a_1a_2c}^{\nu_0\nu_1\nu_2+}$ coincides with (18) up to the
terms which are proportional to $k_i^{\nu_i}$
and give a vanishing contribution
for the polarization vectors $e(k_i)$ satisfying the Lorentz condition
$k_i^{\nu_i}e^{\nu_i}(k_i)=0$ . They are gauge invariant and in the right
light-cone gauge with
$e(k_i)=e_\perp(k_i)-p_B\,(e_\perp k_i)/(p_Bk_i)$
the matrix element of $a$ can be written as follows
\begin{eqnarray*}
e^{\nu_0}(k_0)e^{\nu_1}(k_1)e^{\nu_2}(k_2)a^{\nu_0\nu_1\nu_2}=e_\perp^{\nu_0}
e_\perp^{\nu_1}e_\perp^{\nu_2}m^{\nu_0\nu_1\nu_2}\,,
\end{eqnarray*}
where the tensor $m$ has only transversal components and reads
\begin{eqnarray*}
m^{\nu_1\nu_2\nu_3}=\frac{\delta_\perp^{\nu_1\nu_2}}{2}
\left(-\frac{\beta_2}{t}(\frac{1}{t_1}+\frac{1}{s_{12}})
\,k_{1\perp}^{\nu_0}+
\frac{\beta_1}{s_{12}t}\,k_{2\perp}^{\nu_0}\right)+
\end{eqnarray*}
\begin{eqnarray*}
+\frac{\delta_\perp^{\nu_0\nu_1}}{2}
\left(\frac{\beta_1}{\beta_2}\frac{1}{t_1t}\,
k_{2\perp}^{\nu_2}+\frac{1}{t}(\frac{1}{t_1}+
\frac{1}{s_{12}})(k_{1\perp}^{\nu_2}-\frac{\beta_1}{\beta_2}
k_{2\perp}^{\nu_2})\right)+
\end{eqnarray*}
\begin{eqnarray*}
+\frac{\delta_\perp^{\nu_0\nu_2}}{2}
\left(\frac{\beta_2}{\beta_1}\frac{1}{t_1t}\,
k_{1\perp}^{\nu_1}-\frac{1}{s_{12}t}(k_{2\perp}^{\nu_1}-
\frac{\beta_2}{\beta_1}k_{1\perp}^{\nu_1})\right)\,.
\end{eqnarray*}
Here we took into account that $k_{0\perp}=0$. This expression can be used
for finding the one-loop correction to the residue of the BFKL pomeron.

Let us consider now the quasi-multi-Regge process of the double gluon
emission in
the central rapidity interval at  high energy gluon collisions. The
production amplitude of this process in the Born approximation was
calculated in the second paper of ref. [11] and according to eqs (60),(61)
and (62) from this paper it can be written in the form of above eq. (35) with
\begin{eqnarray*}
\psi_{d_1d_2c_2c_1}^{\nu_1\nu_2+-}=2\,g^2\,\{T_{d_1d}^{c_1}T_{d_2d}^{c_2}
A^{\nu_1\nu_2}+([d_1,\nu_1, k_1] \leftrightarrow [d_2,\nu_2, k_2])\}\,\,,
\end{eqnarray*}
where
\begin{eqnarray*}
A^{\nu_1\nu_2}=-\frac{a_1^{\nu_1}a_2^{\nu_2}}{t}+
\frac{b_1^{\nu_1}b_2^{\nu_2}}{s}\left(1+\frac{s\alpha_1\beta_2}{t}\right)+
\frac{b_1^{\nu_1}c_2^{\nu_2}}{s}\left(\frac{t_2}{s\beta_2(\beta_1+\beta_2)}-
\frac{s\alpha_1\alpha_2}{t}\right)+
\end{eqnarray*}
\begin{eqnarray*}
+\frac{c_1^{\nu_1}b_2^{\nu_2}}{s}
\left(\frac{t_1}{s\alpha_1(\alpha_1+\alpha_2)}-
\frac{s\beta_2\beta_1}{t}\right)-\frac{c_1^{\nu_1}c_2^{\nu_2}}{s}\left(1+
\frac{\kappa}{t}-\frac{s\alpha_2\beta_1}{t}\right)-
\end{eqnarray*}
\begin{eqnarray*}
-2(\delta^{\nu_1\nu_2}-\frac{2k_2^{\nu_1}k_1^{\nu_2}}{\kappa})
\left(1+\frac{t}{\kappa}+\frac{s\alpha_1\beta_2}{t}+
\frac{s\alpha_1\beta_2-s\alpha_2\beta_1}{\kappa}-
\frac{t_1\alpha_2}{\kappa (\alpha_1+\alpha_2)}-
\frac{t_2\beta_1}{\kappa (\beta_1+\beta_2)}\right).
\end{eqnarray*}
Here we introduced the notations
\begin{eqnarray*}
a_1=2\left(\alpha_1p_B+q_1-(\beta_1+\frac{t_1}{s\alpha_1})p_A+
\frac{t}{\kappa}k_2\right),\,
\end{eqnarray*}
\begin{eqnarray*}
a_2=2\left(\beta_2p_A-q_2-(\alpha_2+
\frac{t_2}{s\beta_2})p_B+\frac{t}{\kappa}k_1\right),
\end{eqnarray*}
\begin{eqnarray*}
b_1=2(p_B-\frac{\beta_1s}{\kappa}k_2)\,,\,b_2=2(p_A-\frac{\alpha_2s}{\kappa}
k_1)\,,\,
\end{eqnarray*}
\begin{eqnarray*}
c_1=2(p_A-\frac{\alpha_1s}{\kappa}k_2)\,,\,c_2=2(p_B-
\frac{\beta_2s}{\kappa}k_1)\,,
\end{eqnarray*}
\begin{eqnarray*}
\kappa =(k_1+k_2)^2\,\,,\,\,t=(q_1-k_1)^2\,,\,\,
k_i=\beta_ip_A+\alpha_ip_B+k_{i\perp}\,
\end{eqnarray*}
and $\,q_1=p_A-p_{A'}\,,\,q_2=p_{B'}-p_B\,$.

Note, that due to the gauge invariance $k_1^{\nu_1}A^{\nu_1\nu_2}=0$ of
the amplitude $A^{\nu_1\nu_2}$ the contribution of
the Faddeev-Popov ghosts to the production cross-section is zero [11].
One can verify that the above expression for
$\psi_{d_1d_2c_2c_1}^{\nu_1\nu_2+-}$ coincides with eq. (41) in section 3 up
to the terms proportional to $k_1^{\nu_1}$ or $k_2^{\nu_2}$ which give
a vanishing contribution to the amplitude due to the Lorentz condition
$k_i^{\sigma}e_\sigma(k_i)=0$ for the polarization vectors $e(k_i)$ of the
produced gluons. It is a consequence of the fact, that up to the same
vanishing terms the tensor
$A^{\nu_1\nu_2}$ can be written as the sum of contributions of the
Feynman diagrams with effective vertices given in sections 2 and 3 :
\begin{eqnarray*}
A^{\nu_1\nu_2}=-\frac{1}{2}\frac{\Gamma^{\nu_1\sigma
-}(k_1,k_1-q_1)\Gamma^{\nu_2\sigma
+}(k_2,k_2+q_2)}{(q_1-k_1)^2}-\frac{1}{2}\frac{\gamma^{\nu_2\nu_1\sigma}(k_2,-k_1)
\Gamma^{\sigma +-}(q_2,q_1)}{(k_1+k_2)^2}+
\end{eqnarray*}
\begin{eqnarray*}
+\delta^{\nu_1+}\delta^{\nu_2-}-
\delta^{\nu_1\nu_2}-\frac{1}{2}\delta^{\nu_2+}\delta^{\nu_1-}+4\,t_1\frac{p_A^{\nu_1}
p_A^{\nu_2}}{s^2\alpha_1(\alpha_1+\alpha_2)}+4\,t_2
\frac{p_B^{\nu_1}p_B^{\nu_2}}{s^2\beta_2 (\beta_1+\beta_2)}\,\,\,,
\end{eqnarray*}
where
\begin{eqnarray*}
\Gamma^{\nu_1\sigma -}(k_1,k_1-q_1)=\gamma^{\nu_1\sigma -}(k_1,k_1-q_1)-
t_1n^{-\nu_1}
\frac{1}{k_1^-}n^{-\sigma}\,\,,
\end{eqnarray*}
\begin{eqnarray*}
\Gamma^{\sigma+-}(q_2,q_1)=
\gamma^{\sigma+-}(q_2,q_1)-
2t_1\frac{n^-}{k_1^-+k_2^-}+2t_2\frac{n^+}{k_1^++k_2^+}
\end{eqnarray*}
and $\gamma^{\nu'\nu\sigma}(p_{A'},p_A)$ is the usual Yang-Mills vertex (11).

Because $A^{\nu_1\nu_2}$ satisfies the  Ward identity
\begin{eqnarray*}
k_1^\sigma A^{\sigma \nu_2}=k_2^{\nu_2}\left(\,\frac{1}{2}\,
\frac{k_1^-k_2^+}{(q_1-k_1)^2}-\frac{1}{2}\,\frac{k_1^\sigma
\Gamma^{\sigma +-}(q_2,q_1)}{(k_1+k_2)^2}+1\,   \right)
\end{eqnarray*}
the contribution of the Faddeev-Popov ghosts to the cross-section is fixed.
Instead one can calculate the matrix element of $A^{\nu_1\nu_2}$ between
polarization vectors $e(k_i)$ with definite helicities and sum its
square over all possible helicity states.  It is convenient to
use for $e(k_1)$ and $e(k_2)$ the left and right gauges correspondingly
\begin{eqnarray*}
e^{l}(k_1)=e_\perp^l-\frac{e_\perp^l k_1}{p_A k_1}\,p_A\,,\,\,e^{r}(k_2)=
e_\perp^r-\frac{e_\perp^r k_2}{p_B k_2}\,p_B\,.
\end{eqnarray*}
In these gauges the matrix element of the tensor $A^{\nu_1\nu_2}$ can be
written as follows
\begin{eqnarray*}
e_{\nu_1}^{l\,*}(k_1)\,e_{\nu_2}^{r\,*}(k_2)\,A^{\nu_1\nu_2}=e_{\perp
\nu_1}^{l\,*}(k_1)\, e_{\perp \nu_2}^{r\,*}(k_2)\,a^{\nu_1\nu_2}\,.
\end{eqnarray*}
Here the tensor $a^{\nu_1\nu_2}$ has only transverse components and equals
\begin{eqnarray*}
a^{\nu_1\nu_2}(q_1,q_2;k_1,k_2)=4\{\frac{q_\perp^{\nu_1}q_\perp^{\nu_2}}{t}-
\frac{q_\perp^{\nu_1}}{\kappa}(k_{1\perp}^{\nu_2}-
\frac{\beta_1}{\beta_2}k_{2\perp}^{\nu_2})+\frac{q_\perp^{\nu_2}}{\kappa}
(k_{2\perp}^{\nu_1}-\frac{\alpha_2}{\alpha_1}k_{1\perp}^{\nu_1})+
\end{eqnarray*}
\begin{eqnarray*}
+\frac{k_{1\perp}^{\nu_1}k_{1\perp}^{\nu_2}}{\kappa}
\frac{q_{2\perp}^2}{s\alpha_1(\beta_1+\beta_2)}+
\frac{k_{2\perp}^{\nu_1}k_{2\perp}^{\nu_2}}{\kappa}
\frac{q_{1\perp}^2}{s\beta_2(\alpha_1+\alpha_2)}-
\frac{k_{1\perp}^{\nu_1}k_{2\perp}^{\nu_2}}{\kappa}(1+
\frac{t}{s\alpha_1\beta_2})+
\frac{k_{2\perp}^{\nu_1}k_{1\perp}^{\nu_2}}{\kappa}\}-
\end{eqnarray*}
\begin{eqnarray*}
-2\delta_\perp^{\nu_1\nu_2}(1+\frac{t}{\kappa}+\frac{s\beta_2\alpha_1}{t}+
\frac{s\beta_2\alpha_1-s\alpha_2\beta_1}{\kappa}-\frac{q_{1\perp}^2}{\kappa}
\frac{\alpha_2}{\alpha_1+\alpha_2}-\frac{q_{2\perp}^2}{\kappa}
\frac{\beta_1}{\beta_1+\beta_2}) \,,
\end{eqnarray*}
where $q=q_1-k_1\,,\,t=q^2$. After summing over the polarizations of the
intermediate gluons with the use of the relations
\begin{eqnarray*}
\sum
e_{\sigma'}^{l,r}(k)e_{\sigma}^{l,r\,*}(k)=-\delta_{\sigma \sigma'}^\perp \,,
\,\,\,\,\Omega_{\sigma \sigma'}(k)=\sum e_{\sigma'}^{r}(k)e_\sigma^{l\,*}(k)=
-\delta_{\sigma \sigma'}^{\perp}+2\,
\frac{k_{\perp \sigma}k_{\perp \sigma'}}{k_\perp^2}
\end{eqnarray*}
one can obtain one loop contribution to the BFKL kernel  from the
quasi-multi-Regge kinematics of  intermediate gluons (see [22]). It is
proportional
to the integral over the produced gluon momenta from the quantity :
\begin{eqnarray*}
\sum
A\,A=g_N\,a^{\nu_1\nu_2}(q_1,q_2;k_1,k_2)a^{\nu_1\nu_2}(q_1',q_2';k_1,k_2)+
\end{eqnarray*}
\begin{eqnarray*}
+h_N\,\Omega_{\sigma'\sigma}(k_1)\Omega_{\rho'\rho}(k_2)a^{\sigma \rho}
(q_1,q_2;k_1,k_2)a^{\rho'\sigma'}(q_1',q_2';k_2,k_1)+
(k_1 \leftrightarrow k_2)\,.
\end{eqnarray*}
Here $q_i'=q_i-Q$ and  $Q$ denotes the total momentum transfer
and $g_N$ and $h_N$ are the known
colour factors
\begin{eqnarray*}
g_N=tr \,(T^{i_1})^2(T^{i_2})^2 \,\,,\,\,\,h_N=tr \,(T^{i_1}T^{i_2})^2\,.
\end{eqnarray*}
Matrices $T^i$ are the colour group generators. Note that the
above expression for $\sum A\,A$ is much more complicated, than the result
guessed in ref. [17]. It can not be written as a product of
two factors depending on longitudinal and transverse momenta
correspondingly. The integration over
the squared mass $\kappa$ of the produced gluons contains the ultraviolet
logarithmic divergency . According to our discussion in
section 4 we should introduce an intermediate parameter $\eta$ which allows
us to define  clusters of  particles in the final state. The invariant
mass of the particles inside each cluster is restricted from above by this
parameter. Therefore, after integration over $\kappa$ the result will
contain the term linear in $\eta$ which cancells the analogous infrared
divergency at a
small relative rapidity for two neighbouring gluons produced in the
multi-Regge
kinematics. The finite one-loop contribution to the BFKL kernel
is obtained by the subsequent integration of the constant term
$\sim \eta^0$ over transverse momenta $k_{i\perp}$ for fixed
$q_{1\perp},q_{2\perp}$. This constant term can not be presented as a sum
of contributions  for contracted Feynman diagrams in the transverse subspace
contrary to the assumption made in ref. [17].
Note that in the dispersive
approach developed in ref. [11] one should take into account  the
next to leading corrections to the production amplitude in the
multi-Regge kinematics which also can not be expressed in terms of
the contracted diagrams.



\end{document}